\documentclass[12pt]{article}
\usepackage{amsfonts,amsmath,graphicx,setspace}
\usepackage{color,hyperref,verbatim,natbib,rotating}

\usepackage{booktabs} % to use \toprule, \rulebottom, \cmidrule, \midrule
\usepackage[labelfont=bf]{caption} 
\usepackage{lscape}

\newenvironment{tablenotes}{\list{}{\setlength{\labelsep}{0pt}%
\setlength{\labelwidth}{0pt}%
\setlength{\leftmargin}{0pt}%
\setlength{\rightmargin}{0pt}%
\setlength{\topsep}{2pt}%
\setlength{\itemsep}{0pt}%
\setlength{\partopsep}{0pt}%
\setlength{\listparindent}{0em}%
\setlength{\parsep}{0pt}}%
\item\relax%
}{\endlist}%

\usepackage{xcolor} % to use columns with color
\definecolor{lightgray}{rgb}{0.8, 0.8, 0.8} % to use columns with color
\usepackage{colortbl} % to use columns with color
\newcolumntype{g}{>{\columncolor{lightgray!40}}c} % to use columns with color with color

\definecolor{Red}{rgb}{0.5,0,0}
\definecolor{Blue}{rgb}{0,0,0.5}
\hypersetup{%
    colorlinks = {true},
    linktocpage = {true},
    plainpages = {false},
    linkcolor = {Blue},
    citecolor = {Blue},
    urlcolor = {Red},
    pdfstartview = {XYZ null null 1.25},
    pdfpagemode = {UseOutlines},
    pdfview = {XYZ null null null}
}
\newcommand{\email}[1]{\href{mailto:#1}{\normalfont\texttt{#1}}}

\bibpunct{(}{)}{;}{a}{}{,}

\begin{document}

%{\vspace*{1cm}}

\begin{center}
\singlespacing\Large \bf Efficiently analyzing large patient registries with Bayesian joint models for longitudinal and time-to-event data
\end{center}

%\vspace{0.5cm}

\begin{center}
{\large Pedro Miranda Afonso$^{1,2,\ast}$, Dimitris Rizopoulos$^{1,2}$, Anushka K. Palipana$^{3}$, Grace C. Zhou$^{3,4}$, Cole Brokamp$^{3,5}$, Rhonda D. Szczesniak$^{3,5,6}$ and Eleni-Rosalina Andrinopoulou$^{1,2}$}\footnote{$^\ast$Correspondence at: Department of Biostatistics, Erasmus University Medical Center, PO Box 2040, 3000 CA Rotterdam, the Netherlands. E-mail address: \email{p.mirandaafonso@erasmusmc.nl}.}\\
\singlespacing{
$^{1}$Department of Biostatistics, Erasmus University Medical Center, the Netherlands\\
$^{2}$Department of Epidemiology, Erasmus University Medical Center, the Netherlands\\
$^{3}$Division of Biostatistics and Epidemiology, Cincinnati Children’s Hospital Medical Center, USA\\
$^{4}$Division of Statistics and Data Science, University of Cincinnati, USA\\
$^{5}$Department of Pediatrics, University of Cincinnati, USA\\
$^{6}$Division of Pulmonary Medicine, Cincinnati Children’s Hospital Medical Center, USA\\
}
\end{center}

\vspace{0.5cm}

\begin{spacing}{1}
\noindent {\bf Abstract}\\
The joint modeling of longitudinal and time-to-event outcomes has become a popular tool in follow-up studies. However, fitting Bayesian joint models to large datasets, such as patient registries, can require extended computing times. To speed up sampling, we divided a patient registry dataset into subsamples, analyzed them in parallel, and combined the resulting Markov chain Monte Carlo draws into a consensus distribution. We used a simulation study to investigate how different consensus strategies perform with joint models. In particular, we compared grouping all draws together with using equal- and precision-weighted averages. We considered scenarios reflecting different sample sizes, numbers of data splits, and processor characteristics. Parallelization of the sampling process substantially decreased the time required to run the model. We found that the weighted-average consensus distributions for large sample sizes were nearly identical to the target posterior distribution. The proposed algorithm has been made available in an \textsf{R} package for joint models, \texttt{JMbayes2}. This work was motivated by the clinical interest in investigating the association between ppFEV\textsubscript{1}, a commonly measured marker of lung function, and the risk of lung transplant or death, using data from the US Cystic Fibrosis Foundation Patient Registry (35,153 individuals with 372,366 years of cumulative follow-up). Splitting the registry into five subsamples resulted in an 85\% decrease in computing time, from 9.22 to 1.39 hours. Splitting the data and finding a consensus distribution by precision-weighted averaging proved to be a computationally efficient and robust approach to handling large datasets under the joint modeling framework.\\\\
\noindent {\it Keywords:} big data, consensus Monte Carlo, distributed inference, joint model.
\end{spacing}

%\vspace{0.8cm}

%=====================================================

\section{Introduction}\label{sec:int}

The joint modeling of longitudinal and time-to-event data has become a popular tool in follow-up studies to explore the association between such outcomes~\citep{rizopoulos2012joint}. The availability of large-scale datasets, such as patient registries, is a valuable resource for research to enhance our understanding of disease outcomes and management. However, applying sophisticated statistical models, such as joint models, to large datasets may result in prohibitive computing times. The long running times are particularly cumbersome during the model-building phase, which often requires fitting and comparing multiple models. Different approaches have been described in the body of Bayesian literature to overcome the time-consuming nature of the posterior samplers, such as streamlining the estimation process via analytical approximations, or reducing the amount of data or outcomes analyzed simultaneously. In particular, as an alternative to Markov chain Monte Carlo (MCMC) sampling methods, the model posterior distribution can be approximated with a simpler distribution determined asymptotically. This simplifies the parameter estimation and could therefore allow the analysis of more complex model structures. Examples of such approaches are the Laplace integral approximation~\citep{rue2009approximate} and variational Bayes~\citep{jaakkola2000bayesian} methods. \cite{rustand2022fast} recently presented a joint model for multivariate longitudinal markers and competing risks based on the integrated nested Laplace approximation. \cite{mauff2020joint} describe a corrected two-stage method for fitting a joint model with multiple longitudinal outcomes and a survival outcome. While the two-stage approach substantially decreases the time required to run the model, the time gains are limited by the computation necessary to fit a multivariate model in the first stage. For multivariate joint models with a prohibitive number of outcomes, in subsequent work, \cite{mauff2021pairwise} presented a Bayesian adaptation of the pairwise approach introduced by Fieuws and Verbeke~\citep{fieuws2006pairwise}. The strategy proved less effective under the joint model framework; nevertheless, the results are promising and warrant further investigation. 

Our work is motivated by the analysis of the US Cystic Fibrosis Foundation Patient Registry (CFFPR)~\citep{knapp2016cystic}. Cystic fibrosis (CF) is a severe genetic disease that affects the entire body, but its main symptoms primarily affect the lungs. There have been major improvements in CF outcomes over the past few decades, but it remains a life-limiting condition~\citep{farrell2008guidelines}. The CFFPR contains detailed health-related longitudinal data on US individuals living with CF. It is a large, comprehensive dataset containing annual and encounter-based data on demographics and CF outcomes such as lung function. A commonly measured marker of lung function in CF individuals above six years of age is the percentage of predicted forced expiratory volume in one second (ppFEV\textsubscript{1}). Respiratory failure is the primary cause of death for people with CF, and some patients undergo lung transplantation when the disease progresses quickly. Therefore, there is much clinical interest in investigating the association between CF ppFEV\textsubscript{1} decline and the risk of lung transplant or death using joint models. Previous work on the CFFPR has used smaller samples of the registry data to overcome computational burdens~\citep{andrinopoulou2020integrating}. Many applications with CF data have utilized single-center cohorts~\citep{schluchter2002jointly, piccorelli2012jointly, schluchter2019shared, su2021flexible}. Other applications of CFFPR and UK CF data have considered all available data but relied on simpler joint models~\citep{li2017flexible, taylor2020explaining, andrinopoulou2020multivariate, barrett2015joint}.

In the present study, we explore reducing the computational time required to fit a joint model by tackling the amount of data analyzed simultaneously using consensus Monte Carlo methods. In particular, we split the dataset into independent subsamples, and distribute the posterior sampling between different MCMC samplers that each use one of these subsamples. The multiple sets of posterior samples are combined into a consensus distribution approximating the full posterior distribution. The algorithm is “embarrassingly parallel”~\citep{herlihy2012art} because the MCMC samplers can be executed concurrently without communicating. Thus, when paired with the multi-core processors available in today's computers, which allow the execution of multiple processes in parallel, it can substantially reduce the computing time required to run the model. The advantages of this approach are that it is simple, robust, and invariant to the dimension of the parameter space. The consensus distribution is obtained by averaging the sampled chains across the subposterior distributions. The weights reflect the information in each subsample, estimated from the within-sample Monte Carlo variance. This algorithm relies heavily on the assumption of normality. However, according to the Bernstein--von Mises theorem, the posteriors are approximately Gaussian for large sample sizes. The consensus through averaging has shown promising results when applied to simple regression models~\citep{scott2016bayes}---in which the normality assumption is met for most posteriors---but its performance under more complex settings, such as joint models, has not been evaluated to date. Motivated by the CF application, we examine the association between ppFEV\textsubscript{1} and the risk of lung transplant or death using all available CFFPR data by adapting different consensus strategies in the context of complex methods such as the joint model framework. To make the joint analysis of longitudinal and time-to-event outcomes using large datasets easily applicable by others, we implement the reviewed consensus strategies in an \textsf{R} package for joint models, \verb+JMbayes2+~\citep{jmbayes2}, which is publicly available.

The remainder of this article is organized into four main sections. In Section~\ref{sec:meth}, we present a theoretical introduction to joint models and Bayesian inference and modeling. We describe the consensus Monte Carlo algorithm and three methods to combine MCMC posterior samples, and we discuss their implementation in the \textsf{R} package \verb+JMbayes2+~\citep{jmbayes2}. In Section~\ref{sec:sim}, we explore through a simulation study how different sample sizes, numbers of data splits, and computer processor characteristics affect the performance of these consensus methods under the joint model framework. Section~\ref{sec:app} presents the CFFPR case study that motivates this work, our modeling approach, and the results. In Section~\ref{sec:disc}, we provide concluding remarks and suggest future research directions.

\section{Statistical methods}\label{sec:meth}

\subsection{Joint modeling framework}

To model the longitudinal ppFEV\textsubscript{1} with time to transplantation or death, we rely on the joint modeling of longitudinal and survival data framework~\citep{rizopoulos2012joint}. Let $T^*_i$ denote the true failure time in the event-process for the $i$th individual, $i=1,\dots,n$, and $C_i$ the corresponding independent censoring time. The observed failure time is then $T_i$, with $T_i=\min\left\{T^*_i,C_i\right\}$. The event indicator $\delta_i$ is equal to $1$ if $T_i<C_i$ and $0$ otherwise. We denote the longitudinal marker measured at time $t$ by $y_i(t)$. Joint models assume a full joint distribution of the longitudinal and time-to-event processes $\left[y_i(t),T_i\right]$.

Different factorizations of the joint distribution  have been proposed in the literature~\citep{sousa2011review}. In this work, we focus on the shared-parameter joint models. We assume that the time-to-event and longitudinal processes depend on an unobserved process, defined by random effects $\boldsymbol{b}_i$. The observed processes are assumed independent conditional on the random effects, that is, $\left[y_i(t),T_i\mid\boldsymbol{b}_i\right]=\left[y_i(t)\mid\boldsymbol{b}_i\right]\left[T_i\mid\boldsymbol{b}_i\right]$. This is mathematically convenient but computationally intensive. 

The proposed model takes the form
\begin{equation*} \label{eq:jm}
\begin{cases}
\mathcal{G}\left\{\mu_i(t)\right\}=\eta_i(t)= \boldsymbol x_i^\top(t)\boldsymbol\beta +  \boldsymbol z_i^\top(t)\boldsymbol b_i &\text{longitudinal marker}\\
h_i(t)= h_0(t)\exp \left[  \boldsymbol w_i^\top(t) \boldsymbol\gamma + \mathcal{F}\left\{\eta_i(t), \boldsymbol b_i\right\} \alpha \right] & \text{terminal event}\\
\end{cases},
\end{equation*}
where $i=1, \ldots, n$ represent individuals, $\boldsymbol{b}_i \sim \mathcal{N}\left(\boldsymbol{0}, \boldsymbol\Sigma^{-1}\right)$, and $\boldsymbol\Sigma$ is the covariance matrix. To describe the individual-specific time evolution of the longitudinal marker, we specify a generalized linear mixed model~\citep{pinheiro2006mixed}, with $\boldsymbol x_i(t)$ and $\boldsymbol z_i(t)$ denoting the design vectors for the fixed effects $\boldsymbol{\beta}$ and the random effects $\boldsymbol{b}_i$, respectively. The design vectors may incorporate baseline or time-varying exogenous covariates. The fixed effects describe the average evolution of the marker over time. The random effects account for the individual-specific evolution and the within-individual correlation over time. The expected value of $y_i(t)$ is $\mu_i(t)$, $\mathcal{G}(\cdot)$ is the link function, and $\eta_i(t)$ is the linear predictor.

For the terminal event process, we rely on a proportional hazard risk model to describe the time to the terminal event~\citep{cox2018analysis}. The design vector $\boldsymbol{w}_i(t)$ is the parameter vector of covariates with the corresponding vector of regression coefficients $\boldsymbol{\gamma}$; it can include either baseline or exogenous time-varying covariates. The term $\mathcal{F}\left\{\eta_i\left(t\right),\boldsymbol{b}_i\right\}$ describes the functional form that links the longitudinal and terminal event processes. Different functional forms have been described in the literature, such us underlying value, slope, and cumulative effect~\citep{rizopoulos2012joint, mauff2017extension, andrinopoulou2016bayesian}. The magnitude of the association between the two processes is quantified by $\alpha$.

\subsection{Inference and consensus methods} \label{sec:bayes} \label{sec:cons}

We used Bayesian inference to estimate the joint model parameters from the CFFPR dataset. The posterior distribution for the joint model parameter $\theta$ is defined as 
\begin{equation}\label{eq:bayes}
p(\theta\mid\boldsymbol y, \boldsymbol t, \boldsymbol \delta)=\frac{p(\boldsymbol y, \boldsymbol t, \boldsymbol \delta\mid\theta)p(\theta)}{a}\propto p(\boldsymbol y, \boldsymbol t, \boldsymbol \delta\mid\theta)p(\theta),
\end{equation}
where $p\left(\boldsymbol y,\boldsymbol t,\boldsymbol \delta\mid\theta\right)$ is the likelihood of the sample $\left(\boldsymbol y, \boldsymbol t, \boldsymbol \delta\right)$, $p\left(\theta\right)$ is the prior distribution, and $a$ is a normalizing constant to make the posterior density integrate to one. The posterior estimation is based on Markov chain Monte Carlo (MCMC) sampling methods, such as the the Metropolis--Hastings algorithm. The sampling of high-dimensional individual-specific random effects in shared-parameter joint models makes the sampling process time consuming. More details about Bayesian joint model inference are given by \cite{brown2005flexible} and \cite{ibrahim2001bayesian}. Some methods for Bayesian inference and modeling, and MCMC methods for posterior sampling, have been presented by \cite{gelman2013bayesian}.

MCMC samplers have revolutionized statistical inference, but the analysis of large datasets with high-dimensional parameter spaces remains a major challenge. As datasets grow, the computational cost and time required to run MCMC algorithms increase dramatically, making them impractical for many applications. Motivated by the computational burden of fitting joint models in the CFFPR dataset, we describe three computationally inexpensive consensus strategies under the joint modeling framework, in particular, combining all draws together and averaging the individual draws using two different weighting schemes.

We randomly partition the dataset $(\boldsymbol y, \boldsymbol t, \boldsymbol \delta)$ into S disjoint subsamples $(\boldsymbol y_1, \boldsymbol t_1, \boldsymbol \delta_1),\dots,$ $(\boldsymbol y_S, \boldsymbol t_S, \boldsymbol \delta_S)$. We ensure they are independent by putting together all measurements from the same hierarchy level. For example, for a longitudinal dataset with only two hierarchy levels, measurements from the same individual are placed in a single subsample, one subsample can include multiple individuals, and the subsamples do not overlap. Assuming conditional independence across the subsamples given the parameter of interest $\theta$, the posterior distribution in~\eqref{eq:bayes} can be factorized as
\begin{equation*}\label{eq:prod_dens1}
p(\theta\mid\boldsymbol y,\boldsymbol t,\boldsymbol \delta)\propto p\left(\boldsymbol y,\boldsymbol t,\boldsymbol \delta \mid \theta\right)p(\theta)=\left(\prod^S_{s=1}p(\boldsymbol y_s,\boldsymbol t_s,\boldsymbol\delta_s\mid\theta)\right)p(\theta),
\end{equation*}
where $p\left(\boldsymbol y_s,\boldsymbol t_s,\boldsymbol \delta_s\mid\theta\right)$ is the likelihood of the subsample $\left(\boldsymbol y_s, \boldsymbol t_s, \boldsymbol \delta_s\right)$. When appropriate, the prior information $p(\theta)$ should be scaled to $p(\theta)^{1/S}$ to adjust the prior knowledge to the subsample at hand:
\begin{equation*}\label{eq:prod_dens2}
p(\theta\mid\boldsymbol y,\boldsymbol t,\boldsymbol \delta)\propto\prod^S_{s=1}p(\boldsymbol y_s,\boldsymbol t_s,\boldsymbol\delta_s\mid\theta)p(\theta)^{1/S}=\prod^S_{s=1}p(\theta\mid\boldsymbol y_s,\boldsymbol t_s,\boldsymbol\delta_s).
\end{equation*}
As a result, the full posterior distribution can be seen as the product of the $S$ independent subposterior distributions. Each subsample goes through an independent MCMC simulation (in parallel), each of which generates $D$ draws per chain from the subposterior distribution $p(\theta\mid\boldsymbol y_s,\boldsymbol t_s,\boldsymbol\delta_s)$,
\begin{equation*}\label{eq:mth:draws}
\underset{D\times1}{\boldsymbol\theta^k_s}=\left(\theta^k_{s,1}, \dots, \theta^k_{s,D}\right) \sim p\left(\theta \mid \boldsymbol y_s, \boldsymbol t_s, \boldsymbol \delta_s\right),
\end{equation*}
with $s=1,\ldots,S$, for the Monte Carlo Markov chain $k \in \{1, \dots, K\}$. Here, $\theta^k_{s,1}$ is the first MCMC draw in chain $k$ obtained from the subsample $s$ for the parameter $\theta$. We obtain a consensus posterior distribution by combining the MCMC draws from all subsamples. 

\begin{figure}
\centering
\includegraphics[width=\textwidth]{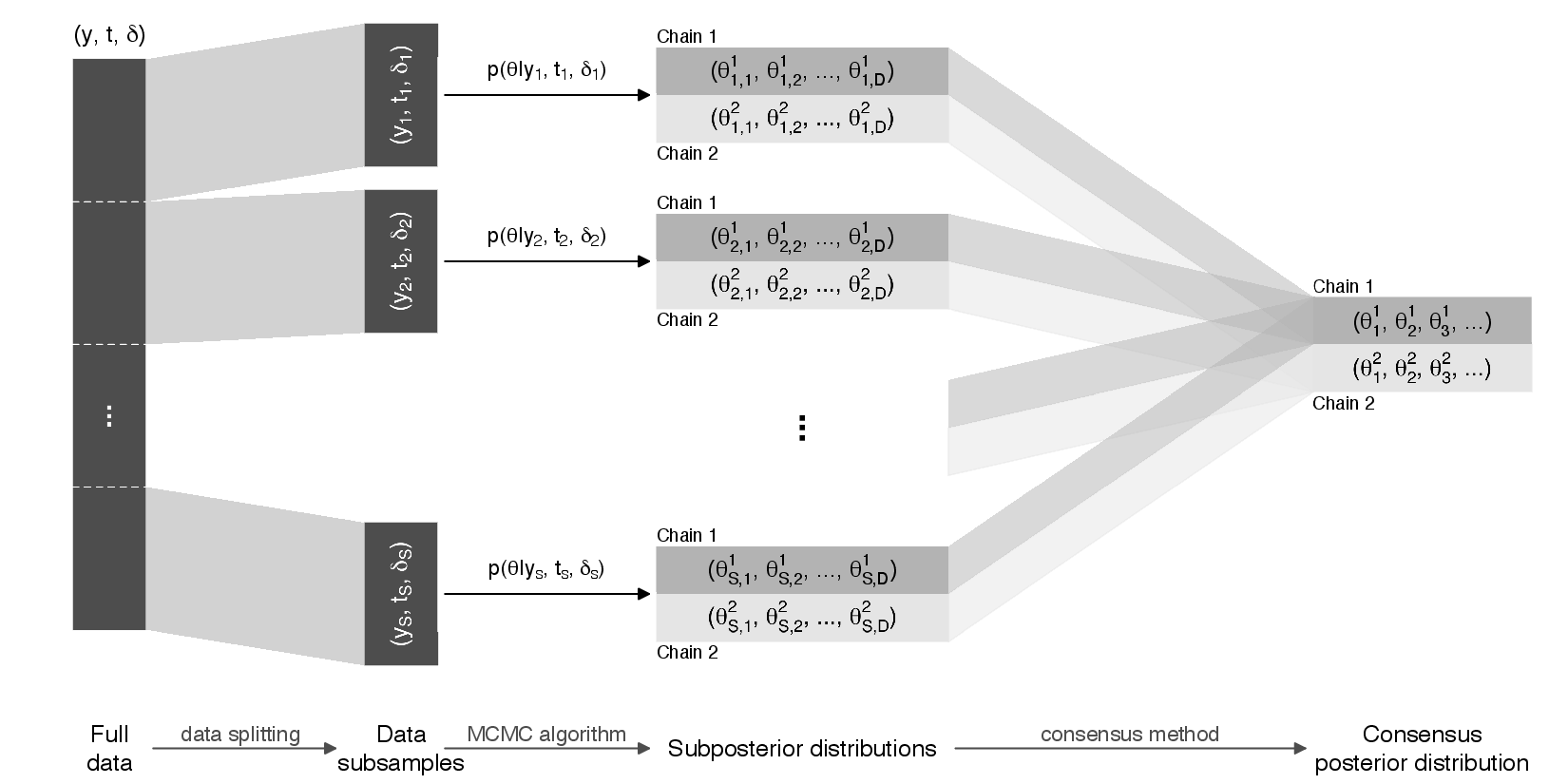}
\caption{Illustration of a consensus Monte Carlo algorithm, considering $S$ data splits with two Markov chains of $D$ draws each.}
\label{fig:mth:dag}
\end{figure}

Figure~\ref{fig:mth:dag} illustrates the consensus Monte Carlo process, considering two Markov chains. The draws from the $k$th chain of each subsample are combined into the $k$th posterior consensus chain. Before the consensus step, the algorithm is embarrassingly parallel, since the MCMC samplers run independently without communicating. This allows the sampling process to be carried out across independent machines to speed up the sampling process. The algorithm is transferable to any MCMC sampler, and the subposterior sampling is carried out in the same manner as sampling from the full-data posterior.

An intuitive approach to producing a consensus posterior distribution is creating a set containing all posterior draws for each chain,
\begin{equation}\label{eq:mth:union}
\underset{(D \times S)\times1}{\boldsymbol\theta^k_{\text{union}}} =(\underset{1\times D}{\boldsymbol\theta^{k\top}_1}, \dots, \underset{1\times D}{\boldsymbol\theta^{k\top}_S}) \sim p(\theta \mid \boldsymbol y, \boldsymbol t, \boldsymbol \delta).
\end{equation}

\newpage

\noindent The number of posterior consensus draws per chain is the product of the combined chains and the number of draws per chain. In the remainder of this article, we refer to this consensus approach as the union algorithm. Even though this approach is straightforward to implement, and the estimates are close to those obtained from the full dataset, it has the disadvantage that it increases the spread of the posterior distribution. To overcome this problem, we can average the individual posterior draws across the $S$ subsamples and combine the $D$ averaged draws. In this case, we have
\begin{equation}\label{eq:mth:wav}
\qquad\underset{D\times1}{\boldsymbol\theta^k_{\text{w-avg}}}=(\overline{\theta_1^k}, \dots, \overline{\theta^k_D}) \sim p(\theta \mid \boldsymbol y, \boldsymbol t, \boldsymbol \delta),
\end{equation}
where 
\begin{equation}\label{eq:wav2}
\overline{\theta^k_d} = \sum_{s=1}^{S} w^k_s\theta^k_{s,d},
\end{equation}
for posterior draw $d \in \{1, \dots, D\}$. The weight applied to the MCMC draws in the $k$th chain from the subsample $s$ is $w_s^k$. The final number of posterior draws equals the number of draws sampled from each subposterior distribution. Compared to the union algorithm, averaging the draws reduces the spread of the consensus posterior distribution around the mean. Table~\ref{tab:mth:cons} illustrates how the individual posterior draws from the different subsamples are combined under the union and weighted-average consensus methods. For illustrative purposes, we consider $D$ posterior draws for each of two Markov chains, and a dataset split into $S$ subsamples. 

One may argue that, for large subsamples, given that the dataset was split randomly, each subposterior sample has roughly the same information. Thus, we can give all subsamples the same relative importance by assigning them equal weights. In this case, each average draw is a simple average or arithmetic mean. Then $w^k_s$ in~\eqref{eq:wav2} becomes
\begin{equation}\label{eq:mth:equal}
w^k_s = \frac{1}{S}\,,
\end{equation}
for each subsample $s$. However, despite the random splitting, the shape of the subposterior densities might still vary between subsamples by chance. We can account for such differences by using information-based weighting to estimate the parameters unbiasedly. \cite{scott2016bayes} proposed a weighting scheme in which the weights reflect the precision in each subposterior sample. Recalling that precision is the inverse of variance, $w_s^k$ in~\eqref{eq:wav2} becomes
\begin{equation}\label{eq:mth:prec}
w^k_s=\frac{w^{k'}_s}{a^k},
\end{equation}
where
\begin{equation}\label{eq:pav2}
w^{k'}_s=\text{Var}^{-1}\boldsymbol\theta^k_s=\text{Var}^{-1}\left(\theta^k_{s,1},\dots, \theta^k_{s,D}\right),
\end{equation}
and $a^k$ is the normalizing constant for the $k$th chain by which all weights sum to one, $a^k=\sum_{s=1}^S w^{k'}_s$. Greater precision leads to a higher weight. Previous research on simpler regression models has demonstrated that the posterior achieved is almost identical to that obtained by fitting all data together~\citep{scott2016bayes}. If the subposterior distributions are normal, that is, $\theta^k_{s,d} \sim p\left(\theta \mid \boldsymbol y_s, \boldsymbol t_s, \boldsymbol \delta_s\right) = \mathcal{N}\left(\mu_s, \tau_s\right)$, then by considering the weight $w^{k'}_s$ as the distribution precision $\tau_s$, the averaged draws are exact random samples from the full-data posterior 
\begin{equation*}
\overline{\theta^k_d}\sim\mathcal{N}\left(\left(\sum_s\tau_s\right)^{-1}\sum_s\tau_s\theta^k_{s,d}, \; \sum_s\tau_s\right)=p\left(\theta \mid \boldsymbol y, \boldsymbol t, \boldsymbol \delta\right),
\end{equation*}
where
\begin{equation*}
\overline{\theta^k_d}=\cfrac{1}{\sum_{s=1}^{S}\tau_s}\;\sum_{s=1}^{S}\tau_s\theta^{k}_{s,d}.
\end{equation*}
Recall that $p_1p_2\sim\mathcal{N}\left(\left(\tau_1+\tau_2\right)^{-1}\left(\tau_1\mu_1+\tau_2\mu_2\right),\tau_1+\tau_2\right)$ when $x_1\sim p_1 =\mathcal{N}\left(\mu_1, \tau_1\right)$ and $x_2\sim p_2 = \left(\mu_2, \tau_2\right)$. The precision $\tau_s$ is unknown, but the sample precision from the Monte Carlo draws, as defined in~\eqref{eq:pav2}, can be used as it is the best estimate $\hat{\tau_s}=\text{Var}^{-1}\boldsymbol\theta^k_s$. Normality is a sufficient condition here but not a necessary one. The Bernstein--von Mises theorem---the Bayesian analogue to the central limit theorem---shows that for large sample sizes the posterior distribution is approximately Gaussian~\citep{van2000asymptotic}. The algorithm's ability to capture characteristics of the joint model posterior distributions has yet to be evaluated.

\subsection{Software and implementation} \label{sec:meth:sfw}

In recent years, joint models have seen considerable software development across primary statistical software tools, such as \textsf{R}, Stata, and SAS. These solutions cover frequentist and Bayesian models and offer a range of model customization options. \cite{furgal2019review} reviewed some of these software implementations. In this work, we implemented the consensus methods described in Subsection~\ref{sec:cons} in \textsf{R} package \verb+JMbayes2+, available from the Comprehensive \textsf{R} Archive Network. \verb+JMbayes2+ is a user-friendly and versatile package that fits Bayesian joint models for longitudinal and time-to-event data~\citep{jmbayes2}.

Today, most computers have multi-core central processing units that enable more efficient processing of multiple tasks simultaneously. However, this does not reduce the computing time by a factor of the number of cores available because there is an overhead due to the indirect computing time used by the operating system to conduct the parallelization process. For fast computation processes, the cost of parallelization may outweigh the benefits of having more processing power, potentially making them take longer. Standard R is single-threaded, but there are packages such as \verb+parallel+ that allow a workload to be split across multiple cores. In our implementation, we concurrently run numerous joint models and various Markov chains to speed up sampling. The package runs parallel independent MCMC simulations for each subsample and returns a consensus distribution, according to the number of splits and the consensus method specified by the user. We run the joint models in parallel, and within each joint model we run its Markov chains in parallel.  One core is required to run each Monte Carlo Markov chain. To maximize the algorithm's efficiency, and thus minimize the computing time, the number of available cores should ideally match the product of the number of splits and the number of Markov chains in each model. However, this is not strictly necessary to apply the techniques and benefit from a faster computation, as further results will show. In the supplementary Section~D, we present an example of the use of consensus methods with \verb+JMbayes2+. 

\section{Simulation study}\label{sec:sim}

We investigated how the union, equal-weighted, and precision-weighted consensus methods perform under the joint model framework through a simulation study. We explored a range of scenarios reflecting different sample sizes, numbers of data splits, and computer processor characteristics. In particular, we considered $n$ individuals, $n\in\left\{500, 1000, 2500, 5000\right\}$, with repeated measurements, and for each dataset we used $s$ data splits, $s\in\left\{1, 2, 5, 10\right\}$. The purpose of the single-split simulations ($s=1$) was to replicate the gold standard approach, in which all data are used together. Furthermore, we investigated two processor architecture scenarios: i) unlimited availability of cores, and ii) 7 cores (to mimic a common 8-core architecture in which one core is set free for other processes outside \textsf{R}). We replicated each scenario 200 times. Table~\ref{tab:sim:gen} outlines the simulation study.
\begin{center}
\begin{table*}% 
\caption{Outline of the simulation study.\label{tab:sim:gen}} 
\small
\begin{tabular*}{0.95\textwidth}{ll}
\toprule 
\textbf{Step} & \textbf{Description} \\
\midrule
1: & Specify the number of individuals $n$, $\text{for } n \in \{500, 1,000, 2,500, 5,000\}$. \\
2: & \quad Repeat 200 times: \\
3: & \quad\quad Simulate dataset from joint model~\eqref{eq:sim:jm}. (Detailed description in Table~\ref{tab:sim:jm}.) \\
4: & \quad\quad Randomly split the $\left(\boldsymbol{y}, \boldsymbol{t}, \boldsymbol{\delta}\right)$ into $S$ subsamples $\left\{\left(\boldsymbol{y}_1, \boldsymbol{t}_1, \boldsymbol{\delta}_1\right), \dots, \left(\boldsymbol{y}_S, \boldsymbol{t}_S, \boldsymbol{\delta}_S\right)\right\}$, \\
   & \quad\quad $\text{for } S \in \{1^\dagger, 2, 5, 10\}$. \\
5: & \quad\quad\quad Specify the number of cores to use $\left\{7,\infty\right\}$.\\
6: & \quad\quad\quad Run $S$ separate MCMC algorithms to sample $\theta_{s,1},\dots,\theta_{s,D}\sim p\left(\theta\mid\boldsymbol y_s, \boldsymbol t_s, \boldsymbol \delta_s\right)$. \\
7: & \quad\quad\quad Combine the resulting MCMC draws according to the consensus methods: \\
   & \quad\quad\quad union~\eqref{eq:mth:union}, equal-~(\ref{eq:mth:wav},~\ref{eq:mth:equal}), and precision-weighted average~(\ref{eq:mth:wav},~\ref{eq:mth:prec}). \\
\midrule
\multicolumn{2}{l}{Datasets count: $\qquad$9,600 $=\underbrace{(\underbrace{4}_{\text{Step 1 }}\times\underbrace{4}_{\text{Step 4} }\times\underbrace{3}_{\text{Step 7} })}_{\text{No. data scenarios}\strut}\underbrace{\times\underbrace{200}_{\text{Step 3}}}_{\text{No. replicas}\strut}$}  \\
\bottomrule 
\end{tabular*} 
\begin{tablenotes} 
\item[] MCMC: Markov chain Monte Carlo. 
\item[$^\dagger$] Full data (gold standard). 
\end{tablenotes} 
\end{table*}
\end{center}

\subsection{Data generation}

We generated data according to the following joint model:
\begin{equation} \label{eq:sim:jm}
\begin{cases}
y_i(t)= \left(\beta_0 + b_{0,i}\right) + \left(\beta_{\text{ns}_1} + b_{\text{ns}_1,i}\right)\text{ns}_{1}(t) + \left(\beta_2 + b_{2,i}\right)\text{ns}_{2}(t) + \left(\beta_3 + b_{3i}\right)\text{ns}_{3}(t) \\
\hphantom{y_i(t)=}+ \varepsilon_i(t) \\
\hphantom{y_i(t)}=  \eta_i(t) + \varepsilon_i(t), \\
h_i(t)= h_0(t)\exp\left\{  \gamma_{\text{sex}} \text{sex}_{\text{male},i} + \gamma_{\text{age}} \text{age}_i + \eta_i(t) \alpha \right\}, \end{cases}
\end{equation}
$t>0$, where $i=1,\dots,n$ represent individuals, $\left(b_{0,i}, b_{\text{ns}_1,i}, b_{\text{ns}_2,i}, b_{\text{ns}_3,i}\right)^\top\sim \mathcal{N}\left(\boldsymbol 0 ,\boldsymbol \Sigma^{-1}\right)$, $\varepsilon_i\sim\mathcal{N}\left(0,\sigma^2_\text{y}\right)$, and $\left(b_{0,i}, b_{\text{ns}_1,i}, b_{\text{ns}_2,i}, b_{\text{ns}_3,i}\right)\perp\!\!\!\perp\varepsilon_i$. The longitudinal marker is described by a linear mixed-effects model. It includes natural cubic splines with three degrees of freedom to accommodate the nonlinear progression of time for both fixed and random effects, represented by $\text{ns}_{j}(t)$, $j=1,2,3$ in~\eqref{eq:sim:jm}. The true level of the longitudinal outcome at time $t$, that is, the observed $y_i(t)$ without the measurement error $\varepsilon_i(t)$, is denoted by $\eta_i(t)$. We assume a Weibull baseline hazard $h_0(t)$ in the proportional hazard risk model, $h_0(t)=\phi\sigma_0t^{\sigma_0-1}$. The relative risk of the terminal event is influenced by continuous and binary time-invariant variables: baseline age and sex, respectively. The parameter values used are in Table~\ref{tab:sim:par}; the characteristics of the simulated datasets are in Table~\ref{tab:sim:data}. We randomly split the individuals from each dataset according to the different strategies. By doing so, we ensure that all subsamples are independent of each other.

\subsection{Model fitting}

The joint models were run using \verb+JMbayes2+ (v0.4-5) on a 64-thread machine\footnote[1]{AMD Ryzen Threadripper PRO 3975WX 32-core 64-thread processor running at 3.49 GHz, using 256 GB of RAM, running Windows 11 Pro (v21H2).}. We restricted the number of available cores to 7 and 63. For the data scenarios considered, the maximum number of cores required at the same time was 30 ($3\text{ chains}\times10\text{ subsamples}$). Thus, 63 cores were enough to reproduce the unlimited number of cores scenario. We considered the true model structure, except for the risk model baseline hazard: we replaced the Weibull baseline risk with penalized B-spline functions~\citep{eilers1996flexible} with 15 knots. For each model, we used the package's default prior distributions and 3 Markov chains with 3500 iterations per chain, discarding the first 500 iterations as a warm-up. The convergence of the chains was assessed by the convergence diagnostic $\hat{R}$~\citep{gelman1992inference} and by visual inspection of posterior traceplots of randomly chosen datasets within each scenario.

\subsection{Results}\label{sec:sim:res}

\begin{figure}
\centering
\includegraphics[width=\textwidth]{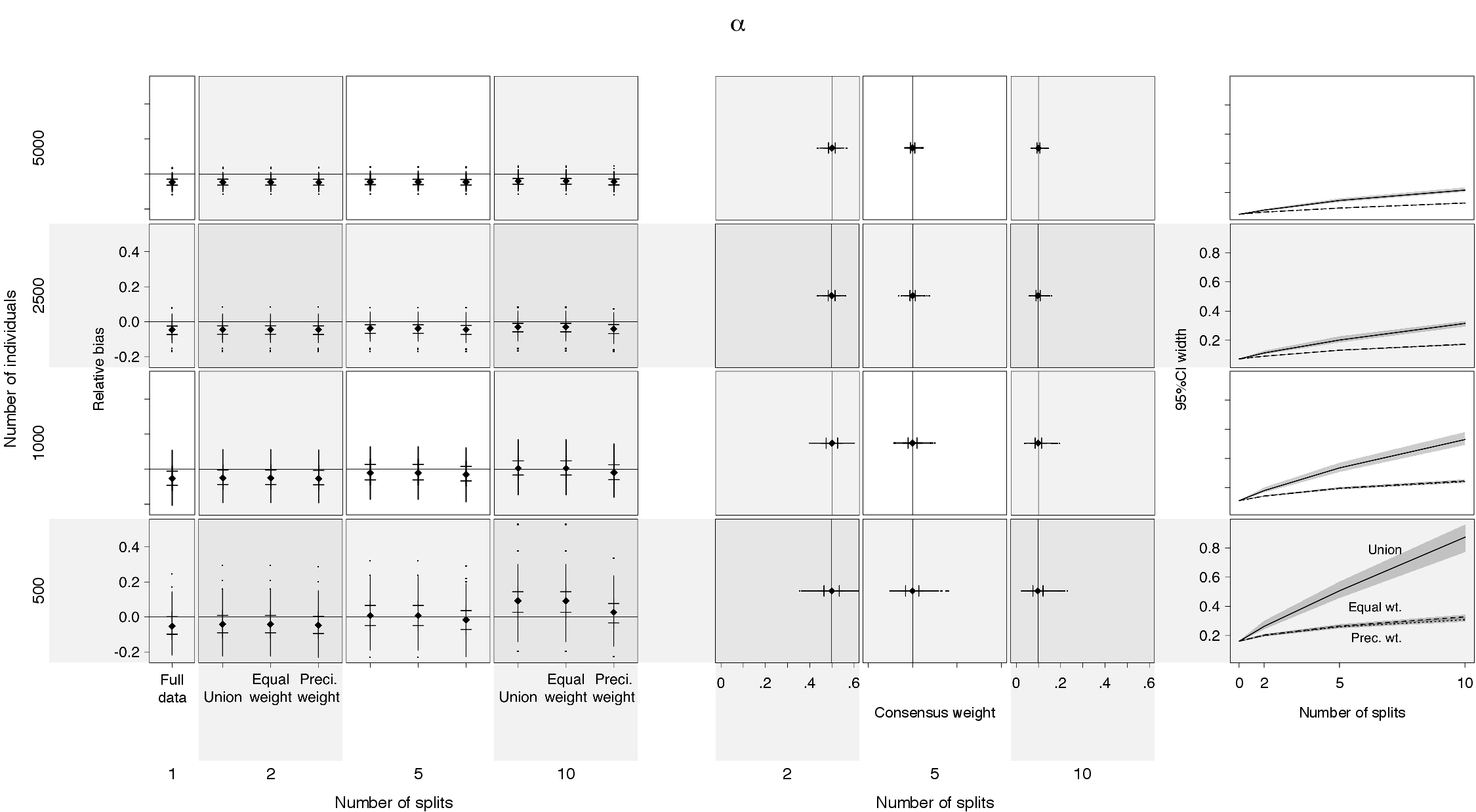}
\caption{Left: Box plot of the relative bias for the $\alpha$ estimate for different sample sizes and numbers of data splits. Center: Consensus (standardized) precision weights for the $\alpha$ estimate for different sample sizes and numbers of data splits. The vertical lines represent the equal weights. Right: Median width of the 95\% credible interval, with the associated interquartile range (IQR), of the $\alpha$ estimate---$1$st, $2$nd, and $3$rd quartiles---against the number of subsamples for different sample sizes.}
\label{fig:sim:mix:alphas}
\end{figure}

There is a general agreement between the full-data (gold standard) estimates and the consensus algorithms' estimates. The relative bias distribution for the association parameter $\alpha$ obtained across the 200 replications is shown in the left panel in Figure~\ref{fig:sim:mix:alphas}, which shows a box plot of relative bias against the consensus algorithm, the sample size, and the number of data splits. Section~B in the supplementary material contains the same plot for the remaining parameters. In general, the consensus methods slightly increase the relative bias compared to the gold standard. A larger sample size reduces the estimation bias. We further notice that when bias is already present in the gold standard approach, minor differences emerge between the different consensus methods, with the precision-weighted consensus yielding a lower value (Figure~\ref{fig:sim:mix:betas4}, left panel, bottom row, and right column).

The three consensus methods accurately capture the location and spread of the posterior distribution of the model parameters. However, the union consensus leads to an over-dispersed distribution. The right panels in Figures~\ref{fig:sim:mix:alphas} and S10--15 illustrate how the 95\% credible interval (CI) width changes as the sample size increases from 50 to 2,500 individuals per subsample. The plot displays the $1$st, $2$nd, and $3$rd quartiles of the 95\% credible interval width against the sample size and the number of data splits. A higher number of data splits, and so less information per subsample, leads to wider credible intervals. However, this effect becomes negligible when the subsamples are sufficiently large. Nonetheless, caution should be taken when estimating effects that are only weakly supported by the data, which may be artifacts introduced by the consensus algorithm. The equal-weighted and precision-weighted algorithms produce similar estimates for the posterior mean and lower and upper percentiles. Given that we split the simulated data randomly, the information asymmetry across subsamples is low. This characteristic makes the precision weights similar across subsamples, approaching the values of the equal weights. The center panels in Figures~\ref{fig:sim:mix:alphas} and \ref{fig:sim:mix:betas1}--\ref{fig:sim:mix:gammas2} present box plots of the standardized consensus precision weights for each model parameter across the 200 replicas against the sample size and the number of data splits. The vertical line represents the equal weight corresponding to the number of data splits. As the size of each subsample increases, the information asymmetry between the subsposterior samples decreases and consequently the precision weights approach the equal weights.

\begin{figure}
\centering
\includegraphics[width=\textwidth]{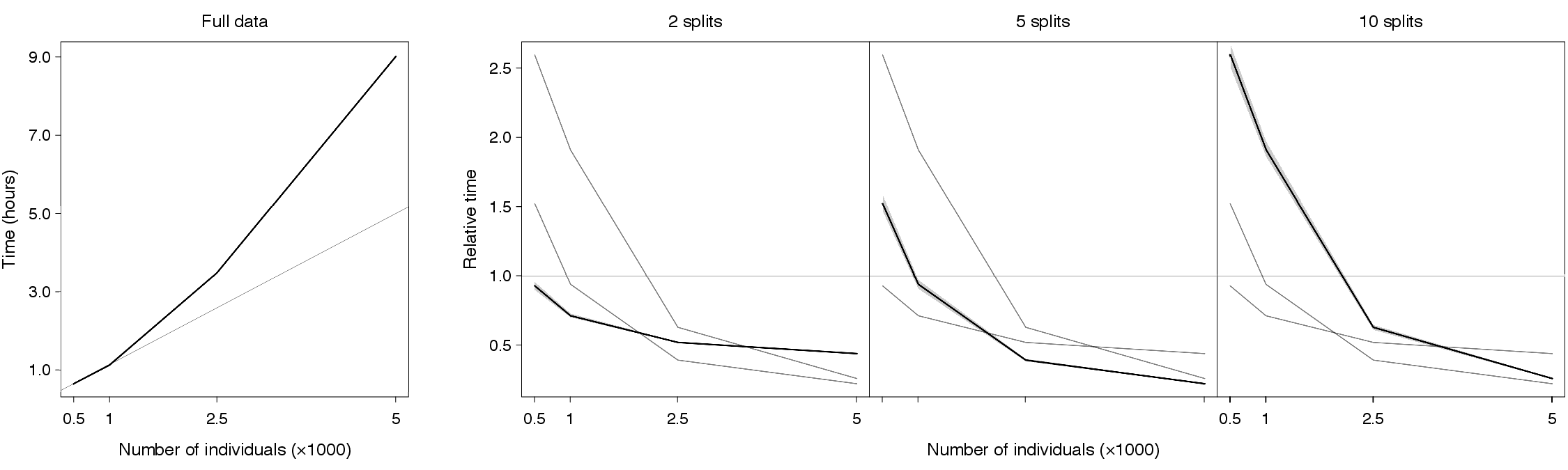}
\caption{Left: Median computing time, with associated IQR, against the number of individuals, when using full data. The gray diagonal line shows a linear evolution. Right: Median and IQR computing time, relative to the time to fit all data together, against the number of individuals, in the scenario with an unlimited number of cores. The gray lines show the median time from the remaining panels.}
\label{fig:sim:time}
\end{figure}

In the left panel in Figure~\ref{fig:sim:time}, we display the computing time required to fit the joint model~\eqref{eq:sim:jm} to datasets of different sizes for the scenario with an unlimited number of cores. As the concave upward curve reveals, the time does not increase linearly with the sample size but instead increases faster for larger sample sizes. Parallel computing comes with overhead costs, which reduce the efficiency of the consensus methods. According to Amdahl's law, the speedup of a process is limited by the fraction of the process that cannot be parallelized, known as the serial fraction~\citep{amdahl1967validity}. In the case of consensus methods, the overhead increases with the number of data splits since more processes need to be created by the operating system, leading to a higher serial fraction. Therefore, the performance of the techniques does not scale linearly as a function of the number of data splits. For example, if the sample size is small, then the overhead may outweigh any advantages of the consensus methods and potentially make the computation take longer. Figure~\ref{fig:sim:time} (right panel) shows the computing time against the sample size, for the three numbers of data splits considered when all required cores were made available. The computing times are presented as proportional increases relative to the median time required to fit all data together. When the relative times are below one, the method is faster than fitting all the data together. The two-split strategy achieves faster computing times than the gold standard across all sample sizes considered. However, the five- and ten-split strategies only become more efficient for the larger sample sizes. Two data splits yield the lowest computing time when $n=$~500, while five data splits produce the lowest computing time when $n=$~1,000, $n=$~2,500, and $n=$~5,000. Although five splits result in the lowest median computing time for $n=$~5,000, the time is similar to that required for ten data splits. The ten-split approach requires higher sample sizes to outweigh its overhead and overcome the five-split strategy. The different splitting strategies allowed us to fit joint models in a timely fashion. For the sample size $n=$~5,000, all splitting strategies reduced the computing time by at least half. Splitting the data into two subsamples reduced the median running time by 56.17\%. The five- and ten-split approaches show a median time reduction of about 0.22 and 0.26, respectively. The absolute computing times are available in Table~\ref{tab:sim:ttime}. The time performance of a given splitting strategy is affected by the number of available cores. If the number of available cores is less than the number of processes to be parallelized, then some processes are placed into a waiting queue, increasing the computing time. The computing times obtained when restricting the number of available cores to seven are available in Figure~\ref{fig:sim:time7} and Table~\ref{tab:sim:ttime7}. In our study, each subsample required three cores available to run its three Markov chains. Thus, the five- and ten-split strategies required more cores simultaneously than the seven made available---15 cores and 30 cores, respectively. The waiting time necessitated by this constraint downgraded the performance of both strategies.

\section{Application} \label{sec:app}
\subsection{The CFFPR dataset}

\begin{figure}
\centering
\includegraphics[width=\textwidth]{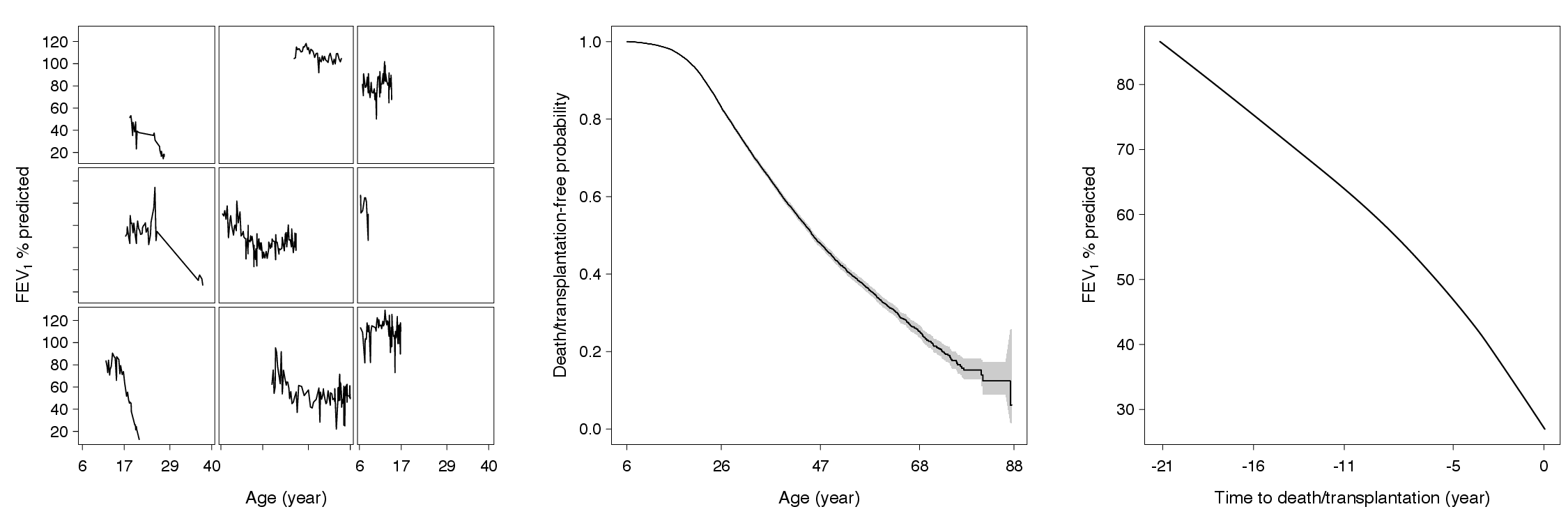}
\caption{Left: ppFEV\textsubscript{1} measurements against age for nine randomly selected individuals. Center: Kaplan--Meier estimate of the death/transplantation-free probability, with associated 95\% confidence interval. Right: ppFEV\textsubscript{1} measurements against time to transplantation or death, with lowess smooth curve.}
\label{fig:eda}
\end{figure}

Our research goals require studying the association between ppFEV\textsubscript{1} and the risk of death or transplantation in CF individuals using the available CFFPR data. This dataset contains health-related information for 35,153 individuals aged over six years, who collectively contributed 1,523,406 observations and 372,366 years of follow-up. The demographic, social, and clinical characteristics of these individuals are displayed in Table~\ref{tab:app:data}. Females account for $48.34\%$ of the group, and the median age at baseline is approximately $8.92$ years (IQR $6.23\mbox{--}18.56$). The median follow-up duration is $10.28$ years (IQR $4.59\mbox{--}16.78$). The dataset encompasses encounters between January 1, 1997, and December 31, 2017, with $50\%$ of encounters between 2005 and 2014. This study focuses on a composite endpoint of death or lung transplantation. During the follow-up period, $23.47\%$ of the individuals experienced one of the two. The median age to die from respiratory failure or to receive a double lung transplant is $27.12$ years (IQR $21.36\mbox{--}36.00$). The median age of the individuals at which their follow-up was censored is $21.33$ years (IQR $14.12\mbox{--}30.94$). Figure~\ref{fig:eda} (center panel) shows the Kaplan--Meier curve for the composite endpoint. Over the total follow-up, the median ppFEV\textsubscript{1} accross all individuals is $73.60$ (IQR $50.30\mbox{--}92.60$). Figure~\ref{fig:eda} (left panel) displays the ppFEV\textsubscript{1} evolution experienced by nine randomly selected individuals over time. Figure~\ref{fig:eda} (right panel) suggests a negative association between ppFEV\textsubscript{1} and the risk of experiencing death/transplantation; it shows the mean ppFEV\textsubscript{1} evolution against the time to death or transplantation, for those who experienced one of the two events. It has been well established that lung function loss in individuals with CF is associated with a worse prognosis~\citep{liou2001predictive}.

\subsection{Analysis}

We assumed a Bayesian joint model of longitudinal and survival data to study the association between ppFEV\textsubscript{1} and the risk of death/transplantation. To reduce the computing time required to evaluate the entire dataset, we split it into subsamples and analyzed them in parallel using the R statistical package \verb+JMbayes2+ (v0.4-5). We then used the union, equal-weighted, and precision-weighted strategies to obtain a consensus posterior. Motivated by our simulation results, we split the sample into five subsamples. By doing so, we expected to substantially reduce the computing time while keeping each subsample sufficiently large to ensure they contained similar amounts of information. We fitted a joint model using all data together to assess the quality of the consensus estimates and the time savings. We term this model fit the "gold standard". This was only possible, however, using a non-conventional computer with sufficient memory. As shown in Table~\ref{tab:app:data}, the five subsamples are similar to each other and are each representative of the full sample. We fitted the following joint model:
\begin{equation} \label{eq:app:jm}
\begin{cases}
\text{ppFEV}_{1\,i}(t) = \left(\beta_0 + b_{0,i}\right) + \left(\beta_{\text{ns}_1} + b_{\text{ns}_1,i}\right)\text{ns}_{1}(t) + \left(\beta_{\text{ns}_2} + b_{\text{ns}_2,i}\right)\text{ns}_{2}(t) + \beta_\text{pa}\text{Pa}_i(t) \\
\hphantom{\text{ppFEV}_{1\,i}(t)=}+ \beta_{\text{ins}}\text{insurance}_{\text{yes},i}(t) + \beta_\text{dpx}\text{dep-idx}_i(t) + \beta_\text{[88,93)}\text{dob}_{{\text{[1988,1993)}},i} \\
\hphantom{\text{ppFEV}_{1\,i}(t)=}+ \beta_\text{[93,98)}\text{dob}_{\text{[1993,1998)},i} + \beta_\text{[98,11]}\text{dob}_{\text{[1998,2011]},i} + \beta_\text{sex}\text{sex}_{\text{male},i} \\
\hphantom{\text{ppFEV}_{1\,i}(t)=}+ \beta_{\text{htz}}\text{F508del}_{\text{htz},i} + \beta_{\text{oth}}\text{F508del}_{\text{other},i} + \varepsilon_i(t) \\
\hphantom{\text{ppFEV}_{1\,i}(t)}= \eta_i(t) + \varepsilon_i(t), \\
h_i(t)=\; h_0(t)\exp\left\{ \gamma_\text{sex}\text{sex}_{\text{male},i} + \eta_i(t) \alpha \right\}, \\
\end{cases}
\end{equation}
where $\left(b_{0,i}, b_{\text{ns}_1,i}, b_{\text{ns}_2,i}\right)^\top\sim \mathcal{N}\left(\boldsymbol 0 ,\boldsymbol \Sigma^{-1}\right)$, $\varepsilon_i\sim\mathcal{N}\left(0,\sigma^2_y\right)$, and $\left(b_{0,i}, b_{\text{ns}_1,i}, b_{\text{ns}_2,i}\right)\perp\!\!\!\perp\varepsilon_i$. We used penalized B-splines functions~\citep{eilers1996flexible} $bs(t)$ to define the baseline hazard $h_0(t)=\exp\{\gamma_{{h_0}\,0}+\sum_{q=1}^{15}\gamma_{{h_0}\,q}bs_q(t)\}$. We assumed a non-linear evolution over time for the linear mixed-effects model, using natural cubic splines with two degrees of freedom, represented by $\text{ns}_{1,i}(t)$ and $\text{ns}_{2,i}(t)$ in~\eqref{eq:app:jm}. The initial time is at age six; for example, at $t=2$
an individual is eight years old. We adjusted the average progression of ppFEV\textsubscript{1} for the following individual baseline characteristics: sex, birth cohort, and genotype. We furthermore assumed time-varying characteristics, such as the presence of infection by \textit{Pseudomonas aeruginosa}, the possession of Medicaid insurance, and the deprivation index, as developed in previous work~\citep{brokamp2019material}. For the random effects structure, we assumed a random intercept and the same non-linear effect of time considered for the fixed effects. Concerning the risk model, we assumed that the risk of death/transplantation is affected by the individual's sex and the underlying value of ppFEV\textsubscript{1}. 

We fitted the models assuming 64 cores\footnote{AMD Ryzen Threadripper PRO 3975WX 32-core 64-thread processor running at 3.49 GHz, using 256 GB of RAM, running Windows 11 Pro (v21H2).}. We considered three Markov chains with 3,500 iterations, of which 500 were discarded for warm-up. The convergence of the chains was assessed by the convergence diagnostic $\hat{R}$~\citep{gelman1992inference} and by visual inspection of the Markov chains' traceplots.

\subsection{Results}

\begin{figure}
\centering
\includegraphics[width=\textwidth]{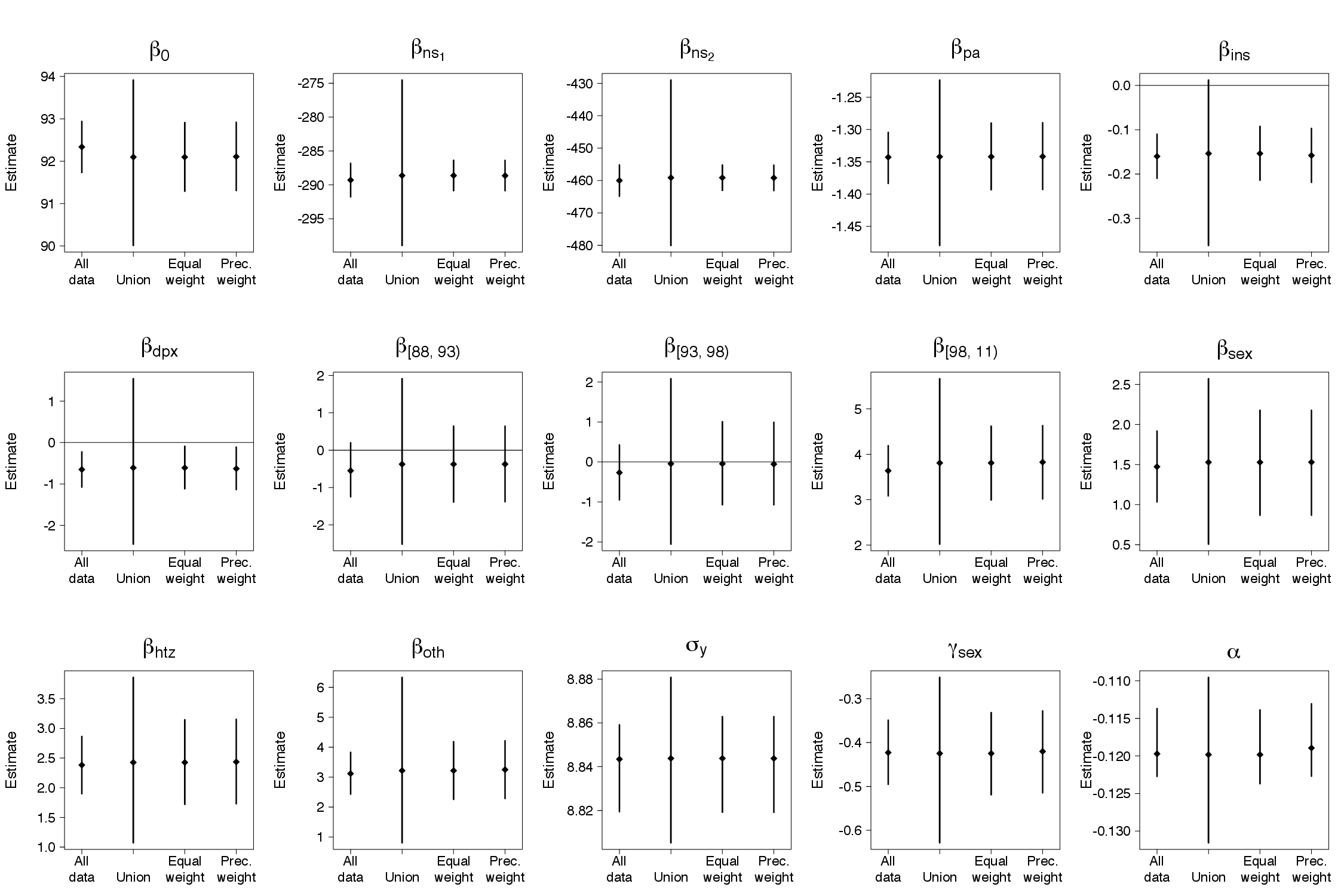}
\caption{Estimated posterior means and 95\% credible interval for joint model coefficients obtained from the full data (gold standard) and different consensus algorithms.}
\label{fig:app:pmean}
\end{figure}

\begin{figure}
\centering
\includegraphics[width=\textwidth]{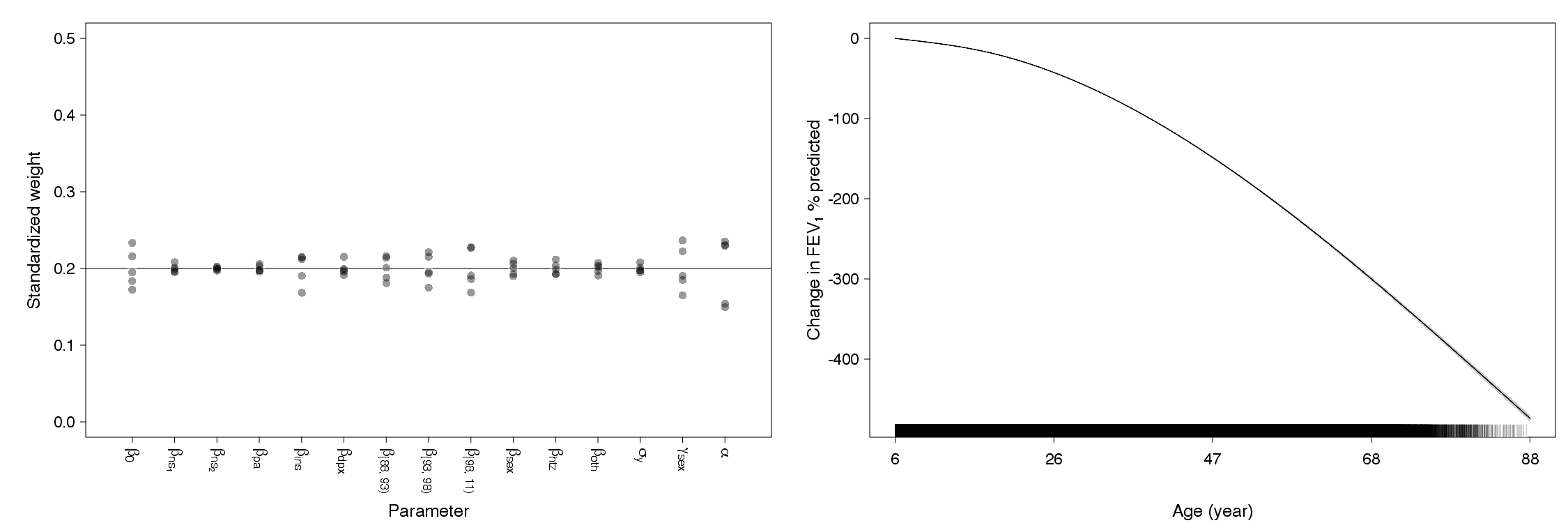}
\caption{Left: Standardized weights used in the precision consensus method for each model parameter; the equal weights are represented by the horizontal line. Right: Main effect of time on the ppFEV\textsubscript{1} progression relative to its initial value at 6 years old.}
\label{fig:app:wts_eff}
\end{figure}

Splitting the full sample, with 35,153 individuals and 1,523,406 samples, into five subsamples produced an $84.89$\% decrease in computing time, from $9.22$ to $1.39$ hours. The results agree with the simulation study findings (Section~\ref{sec:sim:res}). Figure~\ref{fig:app:pmean} shows the estimated posterior means, with associated 95\% credible interval, obtained from the full data (gold standard) and the different consensus methods. The union method consistently led to wider credible intervals. In particular, this method wrongly suggests that the deprivation index and possession of Medicaid insurance have a weak effect on ppFEV\textsubscript{1}, contradicting the findings from the gold standard method. The equal- and precision-weighted approaches performed equally well, yielding posterior estimates close to those obtained when using all data together. Each filled circle in Figure~\ref{fig:app:wts_eff} (left panel) shows the estimated weights for the five different subsamples for that model parameter. The horizontal line represents equal weights of 0.2 each. The precision weights are close to the equal weights, because the random splitting process led to subsamples with similar information, as shown in Table~\ref{tab:app:data}.

The results suggest that ppFEV\textsubscript{1} is negatively associated with the risk of experiencing death or transplantation. The model obtained from the precision-weighted consensus method estimated an association of ${-0.119}$ (95\% CI ${-0.123}$,${-0.113}$), while the model leveraging the entire dataset yielded an estimate of $-0.120$ (95\% CI ${-0.123},{-0.114}$). A 10-unit decrease in the ppFEV\textsubscript{1} increases the hazard by approximately three times ($\text{HR}=3.32$). Table~\ref{tab:app:est} shows the estimates for the remaining parameters obtained using the three consensus methods. For example, the detection of infection by \textit{Pseudomonas aeruginosa} reduces the ppFEV\textsubscript{1} on average by approximately $1.34$. Females present, on average, a lower ppFEV\textsubscript{1} of $1.5$ and a 1.5 times higher hazard of death or transplantation ($\text{HR}=1.521$). Figure~\ref{fig:app:wts_eff} (right panel) displays a main effect plot for time: the average change in ppFEV\textsubscript{1} relative to the ppFEV\textsubscript{1} at baseline against age, considering the gold standard estimates. The concave downward curve indicates that the rate of ppFEV\textsubscript{1} decline increases as the patient gets older. That is, the deterioration of the individual's lung function accelerates with aging.
\begin{sidewaystable}
\caption{Estimated posterior means and 95\% credible interval for the joint model coefficients obtained from the full data (gold standard) and different consensus methods. We present the consensus estimates in terms of differences relative to full-data estimates; the original estimates can be found in Table~\ref{tab:app:est2}.\label{tab:app:est}}
\footnotesize
\begin{tabular*}{\textheight}{@{\extracolsep\fill}lllllllll@{}}
\toprule
& & & \multicolumn{6}{@{}l}{\textbf{Consensus methods}} \\
\cmidrule{4-9}
& \multicolumn{2}{@{}l}{\textbf{Full data}} & \multicolumn{2}{@{}l}{\textbf{Union}} & \multicolumn{2}{@{}l}{\textbf{Equal weight}} & \multicolumn{2}{@{}l}{\textbf{Precision weight}} \\
\cmidrule{2-3} \cmidrule{4-5} \cmidrule{6-7} \cmidrule{8-9}
\textbf{Param.} & \textbf{Mean}  & \textbf{95\% CI}  & $\boldsymbol\Delta$\textbf{Mean}$^{\dagger}$  & $\boldsymbol\Delta$~\textbf{95\% CI}$^{\ddagger}$ & $\boldsymbol\Delta$\textbf{Mean}$^{\dagger}$  & $\boldsymbol\Delta$~\textbf{95\% CI}$^{\ddagger}$ & $\boldsymbol\Delta$\textbf{Mean}$^{\dagger}$  & $\boldsymbol\Delta$~\textbf{95\% CI}$^{\ddagger}$ \\
\midrule
\textbf{LME} & & & \\
$\beta_0$ & $\hphantom{-0}92.337$ & $\hphantom{-0}(91.734, \hphantom{-0}92.939)$ & $\hphantom{-}0.240$ & $\hphantom{0}(1.722,\hphantom{0}{-0.976})$ & ${-0.240}$ & $({-0.441},{-0.026})$ & ${-0.230}$ & $({-0.425},{-0.020})$ \\ 
$\beta_{\text{ns}_1}$ & ${-289.296}$ & $({-291.766}, {-286.828})$ & ${-0.667}$ & $\hphantom{0}(7.182,{-12.297})$ & $\hphantom{-}0.667$ & $\hphantom{-}(0.913,\hphantom{-}0.451)$ & $\hphantom{-}0.652$ & $\hphantom{-}(0.908,\hphantom{-}0.432)$ \\
$\beta_{\text{ns}_2}$ & ${-460.002}$ & $({-464.860}, {-455.208})$ & ${-0.873}$ & $(15.193,{-26.215})$ & $\hphantom{-}0.873$ & $\hphantom{-}(1.798,-0.005)$ & $\hphantom{-}0.821$ & $\hphantom{-}(1.752,{-0.061})$ \\
$\beta_\text{pa}$ & $\hphantom{00}{-1.343}$ & $\hphantom{00}({-1.383}, \hphantom{00}{-1.304})$ & ${-0.001}$ & $\hphantom{0}(0.096,\hphantom{0}{-0.080})$ & $\hphantom{-}0.001$ & $({-0.010},\hphantom{-}0.014)$ & $\hphantom{-}0.001$ & $(-0.010,\hphantom{-}0.014)$ \\ 
$\beta_{\text{ins}}$ & $\hphantom{00}{-0.160}$ & $\hphantom{00}({-0.210}, \hphantom{00}{-0.110})$ & ${-0.006}$ & $\hphantom{0}(0.151,\hphantom{0}{-0.122})$ & $\hphantom{-}0.006$ & $({-0.004},\hphantom{-}0.018)$ & $\hphantom{-}0.002$ & $(-0.009,\hphantom{-}0.013)$ \\
$\beta_\text{dpx}$ & $\hphantom{00}{-0.652}$ & $\hphantom{00}({-1.076}, \hphantom{00}{-0.227})$ & ${-0.041}$ & $\hphantom{0}(1.377,\hphantom{0}{-1.767})$ & $\hphantom{-}0.041$ & $({-0.040},\hphantom{-}0.137)$ & $\hphantom{-}0.020$ & $({-0.060},\hphantom{-}0.115)$ \\ 
$\beta_\text{[88,93)}$ & $\hphantom{00}{-0.548}$ & $\hphantom{00}({-1.247}, \hphantom{-00}0.196)$ & ${-0.172}$ & $\hphantom{0}(1.266,\hphantom{0}{-1.714})$ & $\hphantom{-}0.172$ & $({-0.142},\hphantom{-}0.446)$ & $\hphantom{-}0.175$ & $({-0.133},\hphantom{-}0.444)$ \\ 
$\beta_\text{[93,98)}$ & $\hphantom{00}{-0.268}$ & $\hphantom{00}({-0.947}, \hphantom{-00}0.424)$ & ${-0.223}$ & $\hphantom{0}(1.109,\hphantom{0}{-1.657})$ & $\hphantom{-}0.223$ & $({-0.124},\hphantom{-}0.581)$ & $\hphantom{-}0.214$ & $({-0.125},\hphantom{-}0.567)$ \\ 
$\beta_\text{[98,11]}$ & $\hphantom{-00}3.636$ & $\hphantom{-00}(3.084, \hphantom{-00}4.188)$ & ${-0.173}$ & $\hphantom{0}(1.063,\hphantom{0}{-1.475})$ & $\hphantom{-}0.173$ & $({-0.091},\hphantom{-}0.431)$ & $\hphantom{-}0.190$ & $(-0.067,\hphantom{-}0.441)$ \\
$\beta_\text{sex}$ & $\hphantom{-00}1.472$ & $\hphantom{-00}(1.034, \hphantom{-00}1.917)$ & ${-0.057}$ & $\hphantom{0}(0.527,\hphantom{0}{-0.653})$ & $\hphantom{-}0.057$ & $({-0.166},\hphantom{-}0.260)$ & $\hphantom{-}0.057$ & $({-0.166},\hphantom{-}0.259)$ \\
$\beta_\text{htz}$ & $\hphantom{-00}2.383$ & $\hphantom{-00}(1.900, \hphantom{-00}2.863)$ & ${-0.044}$ & $\hphantom{0}(0.827,\hphantom{0}{-0.996})$ & $\hphantom{-}0.044$ & $({-0.180},\hphantom{-}0.282)$ & $\hphantom{-}0.052$ & $({-0.169},\hphantom{-}0.290)$ \\ 
$\beta_\text{oth}$ & $\hphantom{-00}3.117$ & $\hphantom{-00}(2.431, \hphantom{-00}3.826)$ & ${-0.101}$ & $\hphantom{0}(1.624,\hphantom{0}{-2.504})$ & $\hphantom{-}0.101$ & $({-0.173},\hphantom{-}0.357)$ & $\hphantom{-}0.131$ & $({-0.144},\hphantom{-}0.387)$ \\ 
$\sigma_\text{y}$ & $\hphantom{-00}8.843$ & $\hphantom{-00}(8.819, \hphantom{-00}8.859)$ & ${-0.001}$ & $\hphantom{0}(0.014,\hphantom{0}{-0.022})$ & $\hphantom{-}0.001$ & $\hphantom{-}(0.000,\hphantom{-}0.004)$ & $\hphantom{-}0.001$ & $\hphantom{-}(0.000,\hphantom{-}0.004)$ \\
\textbf{PH} & & & \\
$\gamma_\text{sex}$ & $\hphantom{00}{-0.423}$ & $\hphantom{00}({-0.495}, \hphantom{0}{-0.349})$ & $\hphantom{-}0.002$ & $\hphantom{0}(0.133,\hphantom{0}{-0.097})$ & ${-0.002}$ & $({-0.024},0.017)$ & $\hphantom{-}0.003$ & $(-0.019,\hphantom{-}0.021)$ \\
$\alpha$ & $\hphantom{00}{-0.120}$ & $\hphantom{00}({-0.123}, \hphantom{0}{-0.114})$ & $\hphantom{-}0.000$ & $\hphantom{0}(0.009,\hphantom{0}{-0.004})$ & $\hphantom{-}0.000$ & $({-0.001},0.000)$ & $\hphantom{-}0.001$ & $(\hphantom{-}0.000,\hphantom{-}0.001)$ \\
\bottomrule
\end{tabular*}
\begin{tablenotes}%%[341pt]
\item[] CI: credible interval; LME: linear mixed-effects model; PH: proportional hazards model.
\item[$^\dagger$] $\Delta\text{Mean}=\bar{\theta}_{\text{All data}}-\bar{\theta}_{\text{Consensus method}}$.
\item[$^\ddagger$] $\Delta\text{95\% CI}=\left(P_{0.025}\left\{\boldsymbol\theta_\text{All data}\right\}-P_{0.025}\left\{\boldsymbol\theta_\text{Consensus method}\right\},\, P_{0.9755}\left\{\boldsymbol\theta_\text{All data}\right\}-P_{0.975}\left\{\boldsymbol\theta_\text{Consensus method}\right\}\right)$.
\end{tablenotes}
\end{sidewaystable}

\section{Discussion} \label{sec:disc}

The collection of uniform observational data in patient registries constitutes an important epidemiological research tool that can be used to improve disease outcomes. However, processing such large amounts of data with complex statistical models can be computationally demanding. There is great clinical interest in determining the association between ppFEV\textsubscript{1}, a commonly measured marker of lung function in CF patients, and their risk of death or lung transplantation, using all available CFFPR data. The CFFPR contains health-related information for around 35,000 individuals aged over six, who collectively contributed around 1,500,000 observations. Joint models provide a means to quantify the association between endogenous time-varying covariates and the relative risk of an event of interest~\citep{rizopoulos2012joint}. These models are typically complex as they include multiple submodels with shared random effects. As a consequence, applying them to large datasets can be impractical or even impossible, due to the long running times or memory requirements. A range of approaches have been proposed in the literature to tackle this problem. 

In this work, we studied the embarrassingly parallel consensus Monte Carlo algorithm applied to the joint modeling of longitudinal and survival data framework, with the goal of speeding up the posterior sampling. We randomly divided the dataset into non-overlapping and independent subsamples, used a multi-core processor to analyze them in parallel, and combined the resulting MCMC draws into a consensus distribution. During the sampling process, the MCMC samplers run independently on each subsample without communicating. This characteristic allows the sampling process to be carried out independently in multiple processor cores. This approach can be applied to any MCMC method. 

We explored three consensus strategies---union, equal-weighted, and precision-weighted---that have previously been applied to simpler Bayesian regression models with satisfactory results. These techniques differ in how the posterior draws obtained from the data splits are combined into a single consensus posterior distribution. The union algorithm combines all the MCMC samples. The equal-weighted and precision-weighted algorithms compute a weighted average for each individual posterior draw across the subposterior samples. Through a simulation study, we investigated how these consensus methods perform under the joint model framework for a range of scenarios with different sample sizes, numbers of data splits, and computer processor characteristics. We applied the reviewed methods to the CFFPR case study that motivated this work. We split the CFFPR dataset into five subsamples and compared the computing time and the quality of the estimates against a model that leveraged the entire dataset (the gold standard model). To assist with future longitudinal and time-to-event analyses of large datasets, we implemented the consensus methods in an \textsf{R} package for joint models, \verb+JMbayes2+~\citep{jmbayes2}.

Our simulation results reveal that the computing time required to fit the joint model increases rapidly with the sample size. Furthermore, they show that the consensus Monte Carlo methods can substantially reduce the computing time by parallelizing the sampling process. However, a larger number of data splits does not necessarily translate into reduced computing time. Parallel computing comes with overhead costs, which reduce the efficiency of the consensus methods. For relatively small sample sizes, the overhead associated with a larger number of data splits may outweigh any advantages of the techniques and thus increase the computing time. For example, dividing the data into two splits consistently led to a reduction in computing time for samples of 500, 1,000, 2,500, and 5,000 individuals. However, using five or ten splits only achieved meaningful time savings for sample sizes over 1,000 individuals. These results are sensitive to the complexity of the joint model at hand, such as the number of independent variables and outcomes. More complex models require longer running times; thus, the overhead costs become less significant as the model complexity grows. The efficiency of the methods is also dependent on the processor characteristics. If the number of processes to be parallelized---the number of subsamples multiplied by the number of Markov chains---exceeds the number of available cores, then some of the processes must be placed into a queue, limiting the potential time gains. The number of data splits should therefore be chosen to match the available hardware resources. In our application, splitting the CFFPR into five subsamples yielded an $84.9$\% decrease in computing time, from $9.22$ to $1.39$ hours.

We found an agreement between the posterior mean estimates from the three consensus algorithms and the gold standard model in both the simulation and application studies. The differences lie in the widths of the estimated credibility intervals. The union consensus algorithm is simple to use but generates over-dispersed consensus distributions. This can be problematic in the presence of marginally significant effects. For example, in our application, the union method widened the parameters’ credible intervals, leading to misinterpretation of the relevance of the effect of two of the independent variables in the model. Averaging the draws reduces the spread of the consensus posterior distribution around the mean. The equal-weighted and precision-weighted techniques fared similarly well, yielding posterior samples that were close to those obtained when using the full dataset together. We believe this is because the information asymmetry across the data splits was negligible due to the random splitting of the dataset and the sizes of the subsamples. For large, random subsamples, we expect that the precision-weighted and equal-weighted algorithms should produce similar results. However, in practice, it is difficult to evaluate in advance whether all subsamples are equally informative. Consequently, precision weighting should typically be preferred. 

Our application findings suggest that ppFEV\textsubscript{1} is negatively associated with the risk of dying by respiratory failure or requiring a double lung transplant. That is, the lower the ppFEV\textsubscript{1} value, the higher the risk. While the gold standard approach produced an estimate for this association of $-0.120$ (95\% CI $-0.123,-0.114$), the precision-weighted consensus algorithm estimated an association of $-0.119$ (95\% CI $-0.123$,$-0.113$). The risk is increased by around three times ($\text{HR}=3.32$) for each 10-unit drop in ppFEV\textsubscript{1}. In short, our findings show that our approach accelerates the sampling process while producing correct samples from the target distribution.

This work considered a simple joint model with one longitudinal and one terminal outcome. However, it would be straightforward to extend the reviewed consensus methods to more complex settings, such as joint models accommodating multiple longitudinal outcomes, competing risks, or intermediate, multi-state, or recurrent events. Our goal was to reduce the computing time required to fit a joint model to a large dataset using a single computer; thus, the algorithms presented were described in the context of parallel computation in a single multi-core machine. However, they could be adapted to run across multiple machines in multiple centers to overcome data confidentiality concerns---it would no longer be necessary to share data between centers, but only the model MCMC draws---or sequentially in the same machine to alleviate memory bottlenecks. In this work, we ignored the presence of different demographic groups within the population, that is, we assumed a homogeneous population. When there are large differences between groups, a stratified partition of the dataset might be required. Each subsample must contain enough information to produce unbiased estimates of the parameters: small-sample biases might be introduced if the data are split into many small subsamples. When this is a concern, the bias could be mitigated using, for example, jackknife bias correction~\citep{scott2016bayes}. Further investigation could focus on the impact of the random splitting strategy on the consensus estimates. More specifically, the degree to which the estimated value varies with regard to different groups of random subsamples of the same dataset, to assess the reliability of the various consensus methods. While we expect larger subsample sizes to equalize the variability between the weighted techniques, we anticipate the precision-weighted algorithm to yield a lower variability than the equal-weighted algorithm, since it is robust to asymmetrical subposteriors.

The question of which is the best consensus algorithm to apply to consensus Monte Carlo remains open. In this work, we explored three consensus algorithms. The union algorithm has no theoretical grounding and has major flaws in practice. The precision-weighted algorithm is rooted in Gaussian theory, but given that it does not rely on an explicit Gaussian approximation, it can also capture non-Gaussian features (e.g., skewness, fat tails) of the posterior distribution. Its main weakness is that it is limited to continuous parameter spaces. The equal-weighted consensus can be seen as a relaxation of the precision-weighted algorithm in which, to simplify computation, one assumes that for large sample sizes, the subposterior distributions encapsulate roughly the same information, so their samples can be combined with equal weights. Strategies other than averaging can also be applied to obtain consensus samples. Alternative algorithms have been described in the literature, such as importance resampling~\citep{huang2005sampling}, and modeling the subposterior distributions parametrically~\citep{huang2005sampling}, with kernels~\citep{neiswanger2013asymptotically, wang2013parallel, wang2015parallelizing, srivastava2018scalable}, or with Gaussian process approximations~\citep{nemeth2018merging}. Some of these techniques show promise, but they also add non-trivial complexity. Averaging has powerful advantages: it is simple, stable, and computationally inexpensive. More complex algorithms were not addressed in this work because the simpler techniques produced satisfactory results and the more complex algorithms would increase the computing time. Nonetheless, this should not discourage researchers from exploring other consensus methods under the joint model framework in future work.

Joint models provide a means to estimate individual-specific CF ppFEV\textsubscript{1} decline and its association with mortality or transplantation. The consensus Monte Carlo algorithm is an efficient solution for handling large datasets like the CFFPR without compromising the amount of information taken into account or sacrificing model adequacy, thereby enhancing our understanding of CF ppFEV\textsubscript{1} decline. It could bring new insights into the progression of the disease and could contribute to better monitoring and treatment strategies. The availability of an easy-to-use statistical tool such as \verb+JMbayes2+ is likely to help applied researchers, in the era of big data, perform longitudinal and time-to-event data analyses in their everyday practice.

\subsection*{Acknowledgments}
The authors thank the Cystic Fibrosis Foundation for providing the patient registry (CFFPR) data used to conduct this study, and the patients, care providers, and clinic coordinators at US CF centers for their contributions to the CFFPR.

\subsection*{Funding}
This work was supported by grants from the National Institutes of Health (R01 HL141286) and the Cystic Fibrosis Foundation (SZCZES18AB0).

\subsection*{Data Availability Statement}
The data that support the findings of this study are available from the Cystic Fibrosis Foundation. Restrictions apply to the availability of these data, which were used under license for this study. Requests for data may be sent to datarequests@cff.org.

\newpage

\bibliographystyle{biom}
\bibliography{references}

\setcounter{table}{0} % reset table counter
\renewcommand{\thetable}{S\arabic{table}} % add S to tables label
\setcounter{table}{0} % reset figure counter
\renewcommand{\thefigure}{S\arabic{figure}} % add S to figures label

\begin{landscape}

\section*{Supplementary material}

\appendix

\section{Consensus methods}

\vspace{1.0cm}

\begin{table}[h]
\centering
\caption{Illustration of the union and weighted-average consensus methods, considering $S$ data subsamples with two Markov chains of $D$ draws each.\label{tab:mth:cons}}
\scriptsize
\begin{tabular*}{650pt}{@{\extracolsep\fill}cgcgccgcgcgc@{}}
\toprule
& \multicolumn{7}{@{}l}{\textbf{Subsamples}} & \multicolumn{4}{@{}l}{\textbf{Consensus methods}} \\
\cmidrule{2-8} \cmidrule{9-12}
& \multicolumn{2}{@{}l}{$\boldsymbol{s=1}$} & \multicolumn{2}{@{}l}{$\boldsymbol{s=s}$} & $\boldsymbol\cdots$ & \multicolumn{2}{@{}l}{$\boldsymbol{s=S}$} & \multicolumn{2}{@{}l}{\textbf{Union}} & \multicolumn{2}{@{}l}{\textbf{Weighted average}}\\
\cmidrule{2-3} \cmidrule{4-5} \cmidrule{7-8} \cmidrule{9-10} \cmidrule{11-12}
& \multicolumn{1}{@{}l}{\textbf{Chain 1}} & \multicolumn{1}{@{}l}{\textbf{Chain 2}} & \multicolumn{1}{@{}l}{\textbf{Chain 1}} & \multicolumn{1}{@{}l}{\textbf{Chain 2}} & $\boldsymbol\cdots$ & \multicolumn{1}{@{}l}{\textbf{Chain 1}} & \multicolumn{1}{@{}l}{\textbf{Chain 2}} & \multicolumn{1}{@{}l}{\textbf{Chain 1}} & \multicolumn{1}{@{}l}{\textbf{Chain 2}} & \multicolumn{1}{@{}l}{\textbf{Chain 1}} & \multicolumn{1}{@{}l}{\textbf{Chain 2}} \\
\midrule
$\underset{\textstyle d=1}{\textbf{Draw 1}}$ & $\theta_{1,1}^1$ & $\theta_{1,1}^2$ & $\theta_{2,1}^1$ & $\theta_{2,1}^2$ & $\cdots$ & $\theta_{S,1}^1$ & $\theta_{S,1}^2$ & $\left(\theta_{1,1}^1, \dots, \theta_{S,1}^1\right)^{\top}$ & $\left(\theta_{1,1}^2, \dots, \theta_{S,1}^2\right)^{\top}$ & $\bar{\theta}_1^1=\sum_{s=1}^Sw^1_s\theta_{s,1}^1$ & $\bar{\theta}_1^2=\sum_{s=1}^Sw^2_s\theta_{s,1}^2$ \\ 
$\vdots$ & $\vdots$ & $\vdots$ & $\vdots$ & $\vdots$ & $\ddots$ & $\vdots$ & $\vdots$ & $\vdots$ & $\vdots$ & $\vdots$ & $\vdots$ \\
$\underset{\textstyle d=3}{\textbf{Draw 3}}$ & $\theta_{1,3}^1$ & $\theta_{1,3}^2$ & $\theta_{2,3}^1$ & $\theta_{2,3}^2$ & $\cdots$ & $\theta_{S,3}^1$ & $\theta_{S,3}^2$ & $\left(\theta_{1,3}^1, \dots, \theta_{S,3}^1\right)^{\top}$ & $\left(\theta_{1,3}^2, \dots, \theta_{S,3}^2\right)^{\top}$ & $\bar{\theta}_3^1=\sum_{s=1}^Sw^1_s\theta_{s,3}^1$ & $\bar{\theta}_3^2=\sum_{s=1}^Sw^2_s\theta_{s,3}^2$ \\
$\vdots$ & $\vdots$ & $\vdots$ & $\vdots$ & $\vdots$ & $\ddots$ & $\vdots$ & $\vdots$ & $\vdots$ & $\vdots$ & $\vdots$ & $\vdots$ \\
$\underset{\textstyle d=D}{\textbf{Draw D}}$ & $\underbrace{\theta_{1,D}^1}_{=\underset{D\times1}{\boldsymbol\theta^1_{1}}}$ & $\theta_{1,D}^2$ & $\theta_{2,D}^1$ & $\theta_{2,D}^2$ & $\cdots$ & $\theta_{S,D}^1$ & $\theta_{S,D}^2$ & $\underbrace{\left(\theta_{1,D}^1, \dots, \theta_{S,D}^1\right)^\top}_{\textstyle=\underset{(D \times S)\times1}{\boldsymbol\theta^1_{\text{union}}}}$ & $\left(\theta_{1,D}^2, \dots, \theta_{S,D}^2\right)^{\top}$ & $\underbrace{\bar{\theta}_D^1=\sum_{s=1}^Sw^1_s\theta_{s,D}^1}_{\textstyle=\underset{D\times1}{\boldsymbol\theta^1_{\text{w-avg}}}}$ & $\bar{\theta}_D^2=\sum_{s=1}^Sw^2_s\theta_{s,D}^2$ \\
\bottomrule
\end{tabular*}
\end{table}

\end{landscape}

\section{Simulation study}

\vspace{5cm}

\begin{table*}[h]
\centering %
\caption{Parameter values used in the joint model for data generation in the simulation study.\label{tab:sim:par}}%
\begin{tabular*}{380pt}{lll}
\toprule
\textbf{Model} & \textbf{Parameter}  & \textbf{Value} \\
\midrule
\textbf{Linear mixed-effects} & $\left(\beta_0,\beta_{\text{ns}_1},\beta_{\text{ns}_2},\beta_{\text{ns}_3}\right)$ & $\left(6.94,1.30,1.84,1.82\right)$ \\
& $\sigma_\text{y}^2$ & $0.6^2$ \\
& $\boldsymbol \Sigma$ & $\begin{bmatrix} 0.71 & 0.33 & 0.07 & 1.26 \\  & 2.68 & 3.81 & 4.35 \\ & & 7.62 & 5.40 \\ & & & 8.00 \\\end{bmatrix}$ \\
\textbf{Proportional hazards} & $\phi_0$ & $\exp\left\{-9\right\}$ \\
& $\sigma_0$ & $2$ \\
& $\left(\gamma_{\text{sex}},\gamma_{\text{age}}\right)$ & $\left(0.5,0.05\right)$ \\
& $\alpha$ & $0.5$ \\
\bottomrule
\end{tabular*}
\end{table*}

\begin{center}
\begin{table*} 
\caption{Outline of the joint model data simulation process.\label{tab:sim:jm}} 
\footnotesize
\begin{tabular*}{\textwidth}{@{\extracolsep\fill}ll@{}}
\toprule 
\multicolumn{2}{l}{\textbf{Longitudinal outcome (1/2):}} \\
\midrule
1: & Generate $n$ random samples from $\mathcal{N}\left(\boldsymbol 0 ,\boldsymbol \Sigma^{-1}\right)$ for the individual-specific random effects, $\boldsymbol b_i$: $\underset{n\times3}{\boldsymbol b}$. \\
2: & Generate $(n\times (n_i-1))$ random samples from $\mathcal{U}\left(0,7\right)$ for the visiting times $t_{ij}$, and add the time $0$ \\
& for all individuals: $\underset{(n\cdot n_i)\times1}{\boldsymbol t}$.\\
3: & Generate the B-spline basis vectors for a natural cubic spline with 3 degrees of freedom for the \\
& visiting times $\boldsymbol t$: $\underset{ (n \times n_i)\times4}{\strut\begin{bmatrix} \boldsymbol t_{\text{ns}_1} & \boldsymbol t_{\text{ns}_2} & \boldsymbol t_{\text{ns}_3} \end{bmatrix}}$.
\\
4: & Generate the $n$ vectors of $n_i$ individual underlying longitudinal responses with \\ & $\underset{n_i\times1}{\strut\boldsymbol \eta_i}= \underset{ n_i\times4}{\strut\begin{bmatrix} \boldsymbol 1 & \boldsymbol t_{\text{ns}_1,i} & \boldsymbol t_{\text{ns}_2,i} & \boldsymbol t_{\text{ns}_3,i} \end{bmatrix}}\underset{4\times1}{\strut\boldsymbol\beta} + \underset{n_i\times4}{\begin{bmatrix} \boldsymbol 1 & \boldsymbol t_{\text{ns}_1,i} & \boldsymbol t_{\text{ns}_2,i} & \boldsymbol t_{\text{ns}_3,i} \end{bmatrix}}\underset{4\times1}{\strut\boldsymbol b_i}$. \\
5: & Generate $(n\times n_i)$ random samples from
$\mathcal{N}\left(0,\sigma^2_y\right)$ for the observation measurement error, $\varepsilon_i$: \\ 
& $\underset{(n\cdot n_i)\times1}{\boldsymbol \varepsilon}$.\\
6: & Obtain the observed longitudinal response by summing the vectors $\boldsymbol\eta + \boldsymbol\varepsilon$: $\underset{(n\cdot n_i)\times1}{\boldsymbol y}$.\\
\midrule
\multicolumn{2}{l}{\textbf{Survival outcome:}} \\
\midrule
7: & Generate $n$ random samples from
$\mathcal{U}\left(30,70\right)$ for the individual's baseline age, $\text{age}_i$: $\underset{n\times1}{\textbf{age}}$.\\
8: & Generate $n$ random samples from
$\text{Bern}\left(0.5\right)$ for the individual's sex, $\text{sex}_i$: $\underset{n\times1}{\textbf{sex}}$.\\
9: & Generate $n$ random samples from $\mathcal{U}\left(0, 1\right)$, $u_i$: $\underset{n\times1}{\boldsymbol u}$. \\
10: & Define $H_i(t)=\int_{0}^{t} h_i(t) \,ds$, where $h_i(t)= \underbrace{\phi\sigma_0t^{\sigma_0-1}}_{h_0(t)}\exp\left\{  \gamma_{\text{sex}} \text{sex}_{\text{male},i} + \gamma_{\text{age}} \text{age}_i + \eta_i(t) \alpha \right\}$ \\
11: & Solve numerically $\exp(-H_i(t^*_i))=u_i$ for $t^*_i$~\citep{bender2005generating} to obtain the individual true event \\
& times: $\underset{n\times1}{\textbf{t}^*}$.\\
12: & The observed event times $t_i=\min(t^*_i, t_{\text{max}})$, where $t_{\text{max}}$ is the deterministic maximum follow-up \\ 
& time: $\underset{n\times1}{\textbf{t}}$.\\
13: & Define the type I censoring censoring indicator as $\delta_i=\begin{cases} 1 & t_i \leq t_\text{max}\\ 0 & t_i > t_\text{max} \end{cases}$.\\
\midrule
\multicolumn{2}{l}{\textbf{Longitudinal outcome (2/2):}} \\
\midrule
14: & Remove all $y_{ij}$ for $t_{ij}>t_i$. \\
\bottomrule 
\end{tabular*} 
\end{table*}
\end{center}

\begin{sidewaystable}
\caption{Characteristics of the simulated datasets.\label{tab:sim:data}}%
\scriptsize
\begin{tabular*}{\textheight}{@{\extracolsep\fill}lllll@{\extracolsep\fill}}%
\toprule
\textbf{Number of individuals (n)} & \textbf{500} & \textbf{1,000} & \textbf{2,500} & \textbf{5,000} \\ 
\midrule
\textbf{Number of replicas} & 200 & 200 & 200 & 200 \\
\textbf{Total number of observations} & & & & \\
\multicolumn{1}{r}{Median (IQR)} & 2961.50 (2897.50, 3032.25) & 5927.50 (5849.25, 6026.00) & 14811.50 (14664.75, 14951.50) & 29658.50 (29462.75, 29870.50) \\
\textbf{Number of observations/ind.} & & & & \\
\multicolumn{1}{r}{$2.5^{th}$ PCT, Median (IQR)} & \d1.00 \d(1.00, \d1.00) & \d1.00 \d(1.00, \d1.00) & \d1.00 \d(1.00, \d1.00) & \d1.00 \d(1.00, \d1.00) \\
\multicolumn{1}{r}{Median, Median (IQR)} & \d5.00 \d(5.00, \d5.00) & \d5.00 \d(5.00, \d5.00) & \d5.00 \d(5.00, \d5.00) & \d5.00 \d(5.00, \d5.00) \\
\multicolumn{1}{r}{$97.5^{th}$ PCT, Median (IQR)} & 15.00 (15.00, 15.00) & 15.00 (15.00, 15.00) & 15.00 (15.00, 15.00) & 15.00 (15.00, 15.00) \\
& \raisebox{-\totalheight}{\includegraphics[width=35mm]{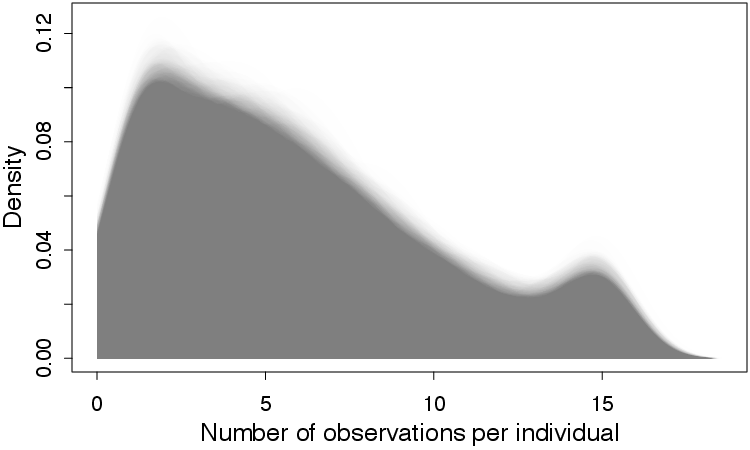}} & \raisebox{-\totalheight}{\includegraphics[width=35mm]{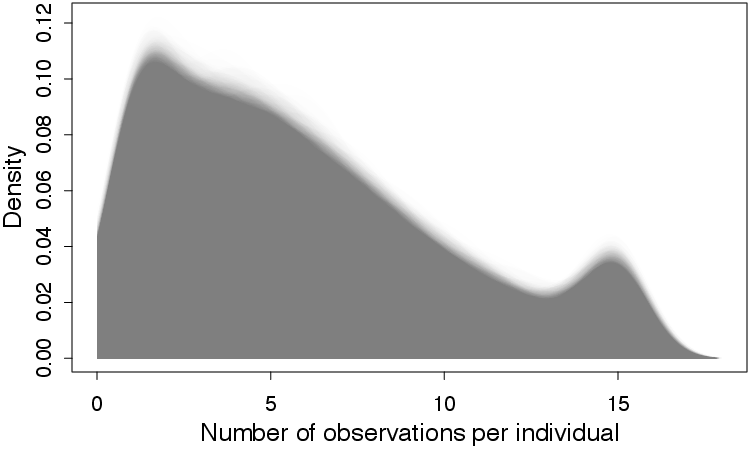}} & \raisebox{-\totalheight}{\includegraphics[width=35mm]{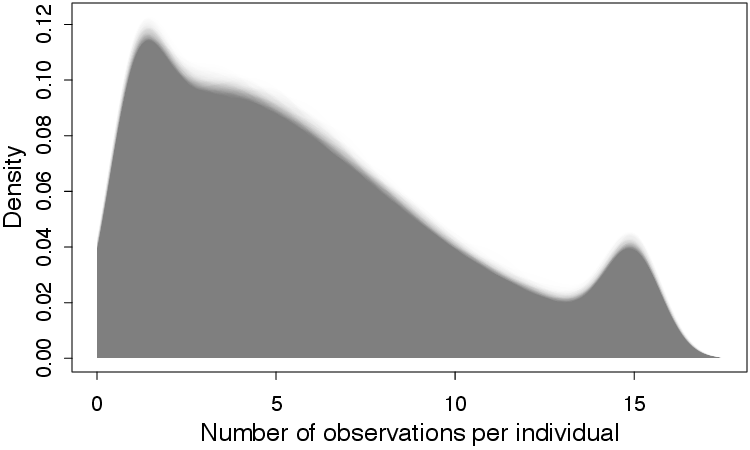}} & \raisebox{-\totalheight}{\includegraphics[width=35mm]{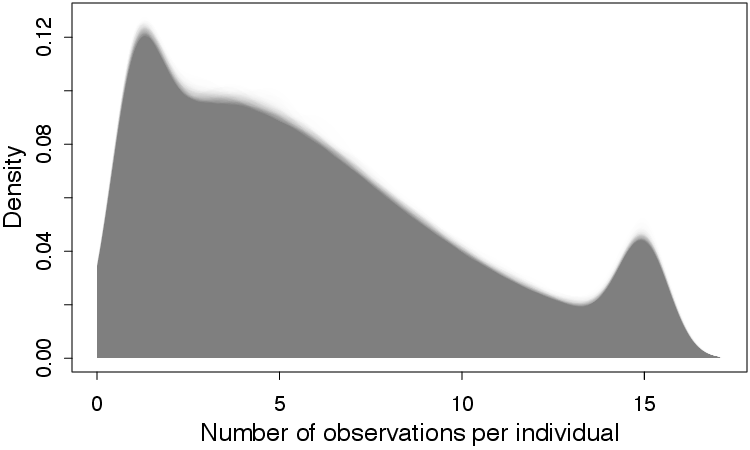}} \\
\textbf{Aggregated follow-up time} & & & & \\
\multicolumn{1}{r}{Median (IQR)} & 1308.38 (1274.51, 1332.12) & 2612.78 (2581.94, 2655.62) & 6524.46 (6460.77, 6585.95) & 13058.13 (12965.20, 13176.31) \\
\textbf{Total follow-up time/ind.} & & & & \\
\multicolumn{1}{r}{$2.5^{th}$ PCT, Median (IQR)} & 0.00 (0.00, 0.00) & 0.00 (0.00, 0.00) & 0.00 (0.00, 0.00) & 0.00 (0.00, 0.00) \\
\multicolumn{1}{r}{Median, Median (IQR)} & 2.38 (2.32, 2.46) & 2.39 (2.35, 2.43) & 2.38 (2.36, 2.41) & 2.39 (2.37, 2.41) \\
\multicolumn{1}{r}{$97.5^{th}$ PCT, Median (IQR)} & 6.77 (6.71, 6.81) & 6.77 (6.73, 6.81) & 6.78 (6.76, 6.80) & 6.78 (6.76, 6.79) \\
& \raisebox{-\totalheight}{\includegraphics[width=35mm]{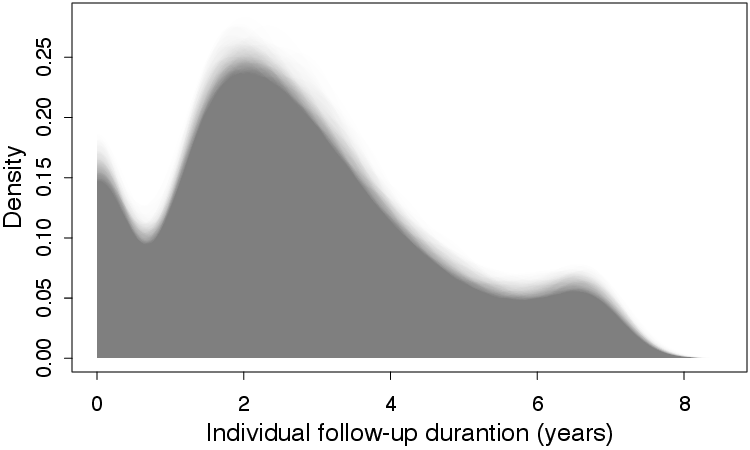}} & \raisebox{-\totalheight}{\includegraphics[width=35mm]{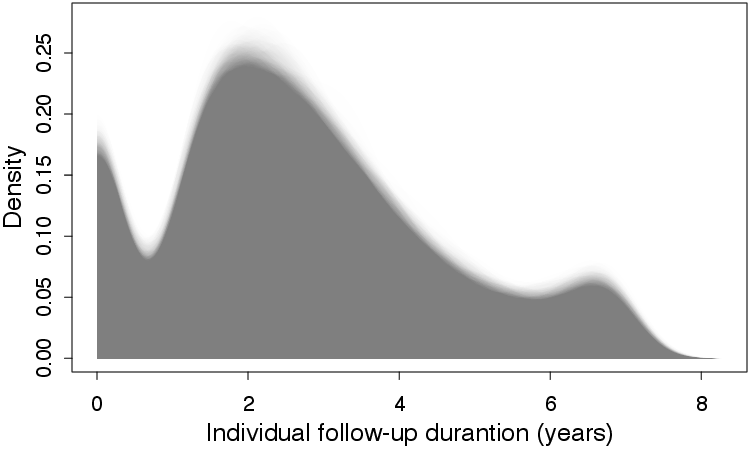}} & \raisebox{-\totalheight}{\includegraphics[width=35mm]{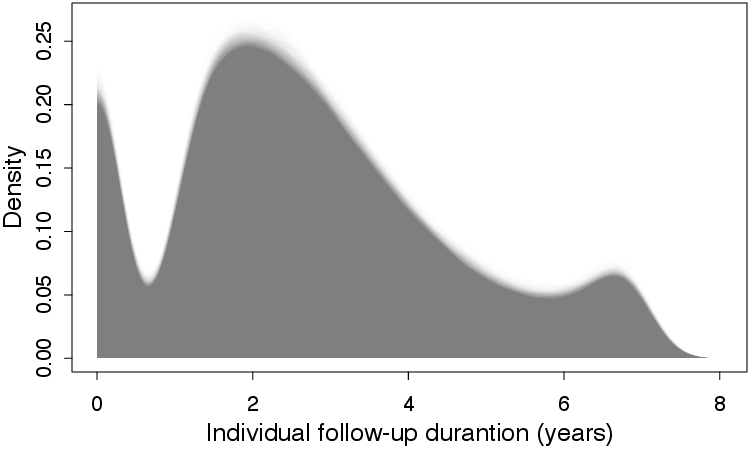}} & \raisebox{-\totalheight}{\includegraphics[width=35mm]{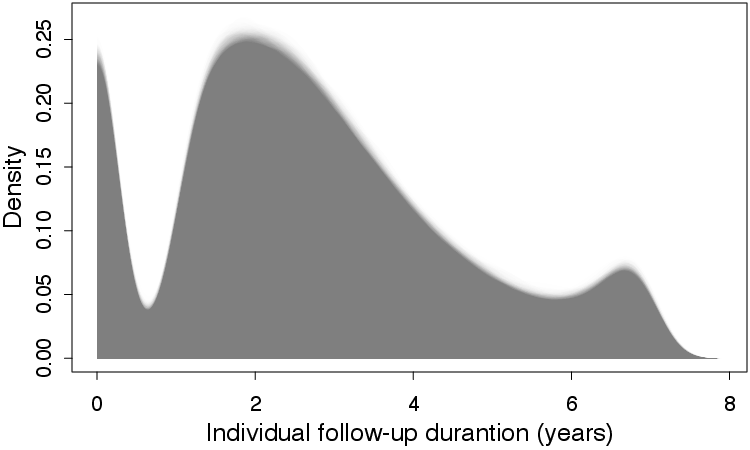}} \\
\textbf{Terminal-event time} & & & & \\
\multicolumn{1}{r}{$2.5^{th}$ PCT, Median (IQR)} & 0.56 (0.50, 0.60) & 0.54 (0.50, 0.58) & 0.54 (0.51, 0.56) & 0.54 (0.52, 0.55) \\
\multicolumn{1}{r}{Median, Median (IQR)} & 2.65 (2.59, 2.71) & 2.65 (2.61, 2.69) & 2.65 (2.63, 2.68) & 2.65 (2.64, 2.68) \\
\multicolumn{1}{r}{$97.5^{th}$ PCT, Median (IQR)} & 6.18 (6.04, 6.30) & 6.19 (6.10, 6.29) & 6.20 (6.15, 6.26) & 6.20 (6.16, 6.23) \\
& \raisebox{-\totalheight}{\includegraphics[width=35mm]{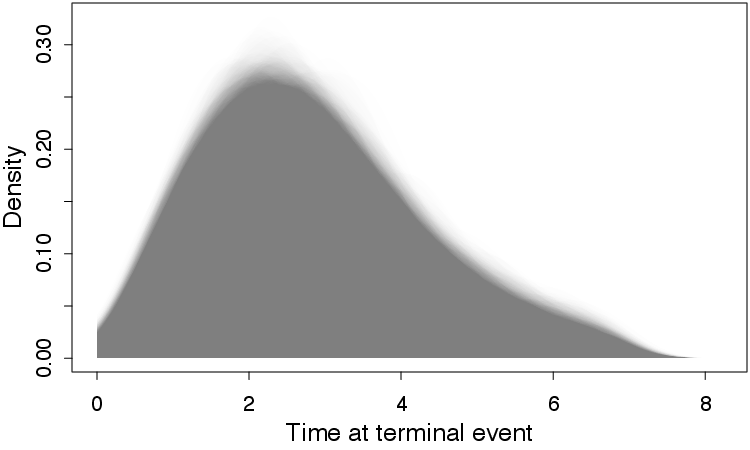}} & \raisebox{-\totalheight}{\includegraphics[width=35mm]{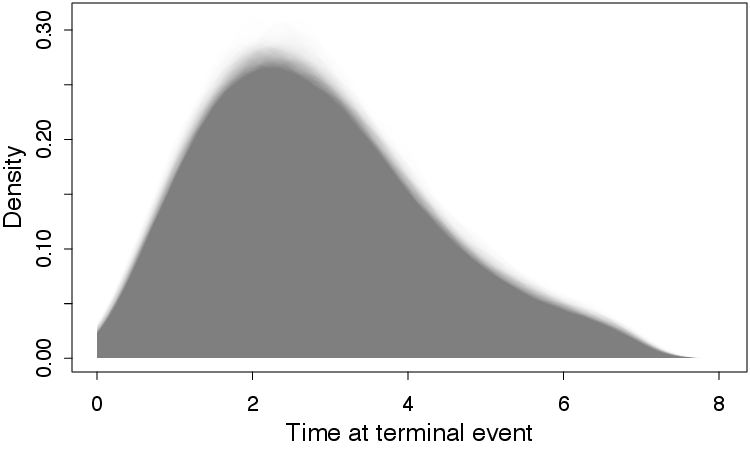}} & \raisebox{-\totalheight}{\includegraphics[width=35mm]{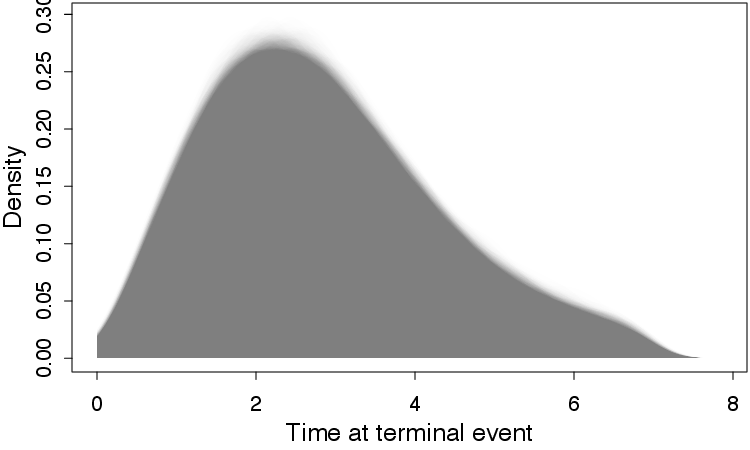}} & \raisebox{-\totalheight}{\includegraphics[width=35mm]{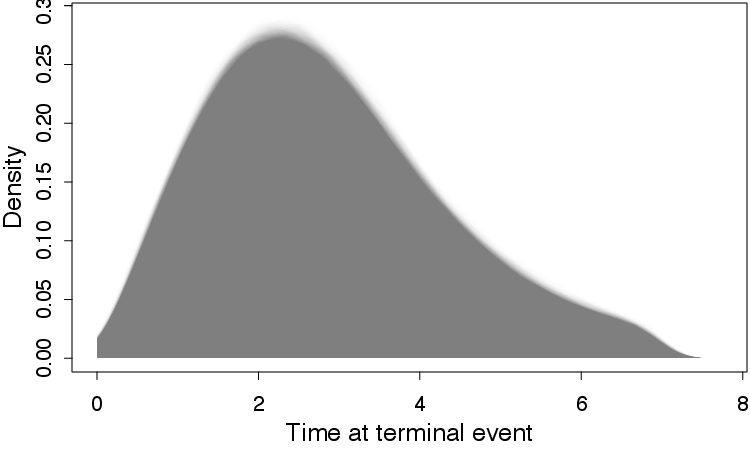}} \\
\textbf{Censoring time} & & & & \\
\multicolumn{1}{r}{Median (IQR)} & 7.00 (7.00, 7.00) & 7.00 (7.00, 7.00) & 7.00 (7.00, 7.00) & 7.00 (7.00, 7.00) \\
\textbf{Terminal event} & & & & \\
\multicolumn{1}{r}{\%, Median (IQR)} & 94.20 (93.40, 94.80) & 94.00 (93.60, 94.60) & 94.12 (93.84, 94.49) & 94.14 (93.90, 94.34) \\
\textbf{Sex (Female)} & & & & \\
\multicolumn{1}{r}{\%, Median (IQR)} & 50.40 (49.60, 51.20) & 50.10 (49.50, 51.00) & 50.34 (49.96, 50.84) & 50.36 (50.04, 50.62) \\
\textbf{Age at baseline} & & & & \\
\multicolumn{1}{r}{Median (IQR)} & 50.14 (49.67, 50.79) & 50.24 (49.74, 50.72) & 50.13 (49.92, 50.47) & 50.23 (50.06, 50.45) \\
\bottomrule
\end{tabular*}
\begin{tablenotes}%%[\textheight]
\item[] IQR: interquartile range; PCT: percentile. 
\end{tablenotes}
\end{sidewaystable}

\begin{landscape}

\begin{table*}
\centering
\caption{Computing time (hours) against the number of individuals and the number of data splits, using 63 cores (200 replicas). The best performance is underlined.\label{tab:sim:ttime}}
\begin{tabular*}{650pt}{@{\extracolsep\fill}lllllllll@{}}
\toprule
& \multicolumn{8}{@{}l}{\textbf{Number of individuals (n)}} \\
\cmidrule{2-9}
& \multicolumn{2}{@{}l}{\textbf{500}} & \multicolumn{2}{@{}l}{\textbf{1,000}} & \multicolumn{2}{@{}l}{\textbf{2,500}} & \multicolumn{2}{@{}l}{\textbf{5,000}} \\
\cmidrule{2-3}\cmidrule{4-5}\cmidrule{6-7}\cmidrule{8-9}
& \textbf{Median} & \textbf{IQR}  & \textbf{Median} & \textbf{IQR} & \textbf{Median} & \textbf{IQR} & \textbf{Median} & \textbf{IQR} \\
\midrule
\textbf{All data}  & 0.655 & (0.648, 0.664)  & 1.137 & (1.123, 1.159) & 3.489 & (3.478, 3.500) & 9.035 & (9.016, 9.051) \\
\textbf{2 splits}  & \underline{0.607} & \underline{(0.587, 0.626)}  & \underline{0.809} & \underline{(0.796, 0.827)} & 1.811 & (1.796, 1.825) & 3.960 & (3.947, 3.975) \\
\textbf{5 splits}  & 0.996 & (0.962, 1.039)  & 1.067 & (1.031, 1.103) & \underline{1.365} & \underline{(1.334, 1.396)} & \underline{1.996} & \underline{(1.957, 2.038)} \\
\textbf{10 splits} & 1.699 & (1.641, 1.748)  & 2.171 & (2.115, 2.239) & 2.191 & (2.126, 2.250) & 2.335 & (2.294, 2.365) \\
\bottomrule
\end{tabular*}
\begin{tablenotes}
\item[] IQR: interquartile range.
\end{tablenotes}
\end{table*}

\begin{table*}
\centering
\caption{Computing time (hours) against the number of individuals and the number of data splits, using 7 cores (200 replicas). The best performance is underlined.\label{tab:sim:ttime7}}
\begin{tabular*}{650pt}{@{\extracolsep\fill}lllllllll@{}}
\toprule
& \multicolumn{8}{@{}l}{\textbf{Number of individuals (n)}} \\
\cmidrule{2-9}
& \multicolumn{2}{@{}l}{\textbf{500}} & \multicolumn{2}{@{}l}{\textbf{1,000}} & \multicolumn{2}{@{}l}{\textbf{2,500}} & \multicolumn{2}{@{}l}{\textbf{5,000}} \\
\cmidrule{2-3}\cmidrule{4-5}\cmidrule{6-7}\cmidrule{8-9}
& \textbf{Median} & \textbf{IQR}  & \textbf{Median} & \textbf{IQR} & \textbf{Median} & \textbf{IQR} & \textbf{Median} & \textbf{IQR} \\
\midrule
\textbf{All data}  & 0.583 & (0.579, 0.589)  & 1.057 & (1.052, 1.116) & 3.426 & (3.342, 3.441) & 8.861 & (8.823, 8.987) \\
\textbf{2 splits}  & \underline{0.438} & \underline{(0.432, 0.442)}  & \underline{0.661} & \underline{(0.656, 0.665)} & \underline{1.647} & \underline{(1.642, 1.652)} & 3.709 & (3.677, 3.802) \\
\textbf{5 splits}  & 0.859 & (0.850, 0.870)  & 1.081 & (1.071, 1.089) & 1.840 & (1.832, 1.853) & 3.384 & (3.366, 3.428) \\
\textbf{10 splits} & 1.265 & (1.257, 1.275)  & 1.473 & (1.460, 1.483) & 2.093 & (2.076, 2.114) & \underline{3.187} & \underline{(3.170, 3.208)} \\
\bottomrule
\end{tabular*}
\begin{tablenotes}
\item[] IQR: interquartile range.
\end{tablenotes}
\end{table*}

\end{landscape}

\begin{figure}
\centering
\includegraphics[width=\textwidth]{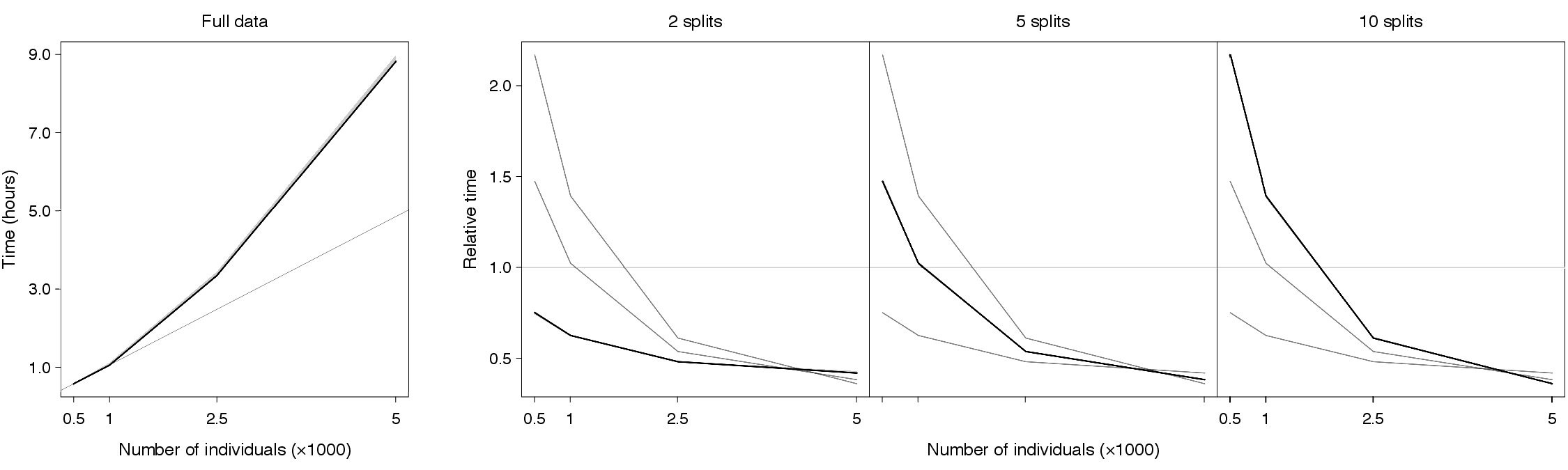}
\caption{Left: Median computing time, with associated IQR, against the number of individuals, when using all data together, using 7 cores. The gray diagonal line shows a linear evolution. Right: Median and IQR computing time, relative to the time required to fit all data together, against the number of individuals, in the scenario with an unlimited number of cores, using 7 cores. The gray lines show the median time from the remaining panels.}
\label{fig:sim:time7}
\end{figure}

\begin{figure}
\centering
\includegraphics[width=\textwidth]{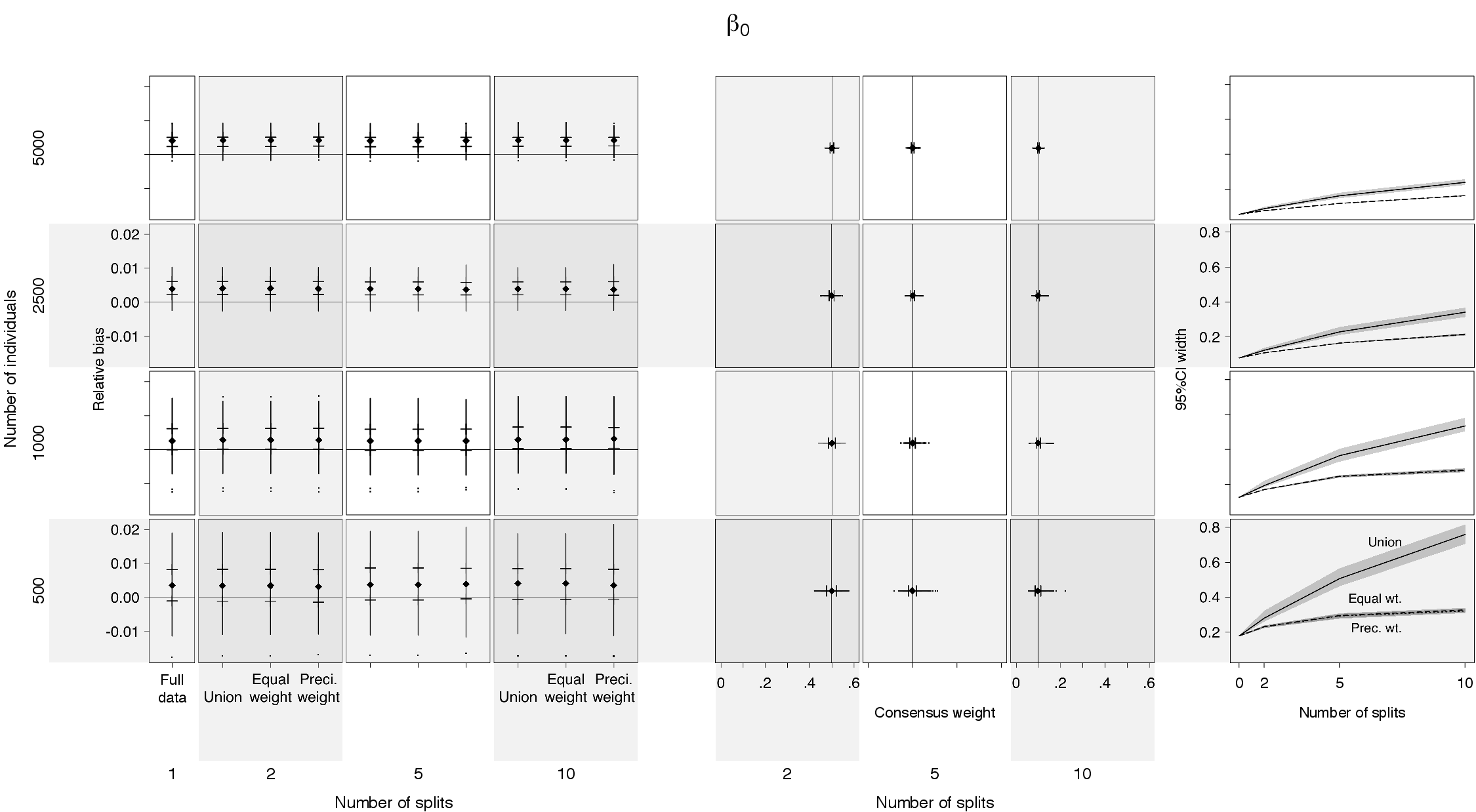}
\caption{Left: Box plot of the relative bias for the $\beta_0$ estimate for different sample sizes and numbers of data splits. Center: Consensus (standardized) precision weights for the $\beta_0$ estimate for different sample sizes and numbers of data splits. The vertical lines represent the equal weights. Right: Median width of the 95\% credible interval, with the associated IQR, of the $\beta_0$ estimate---$1$st, $2$nd, and $3$rd quartiles---against the number of splits for different sample sizes.}
\label{fig:sim:mix:betas1}
\end{figure}

\begin{figure}
\centering
\includegraphics[width=\textwidth]{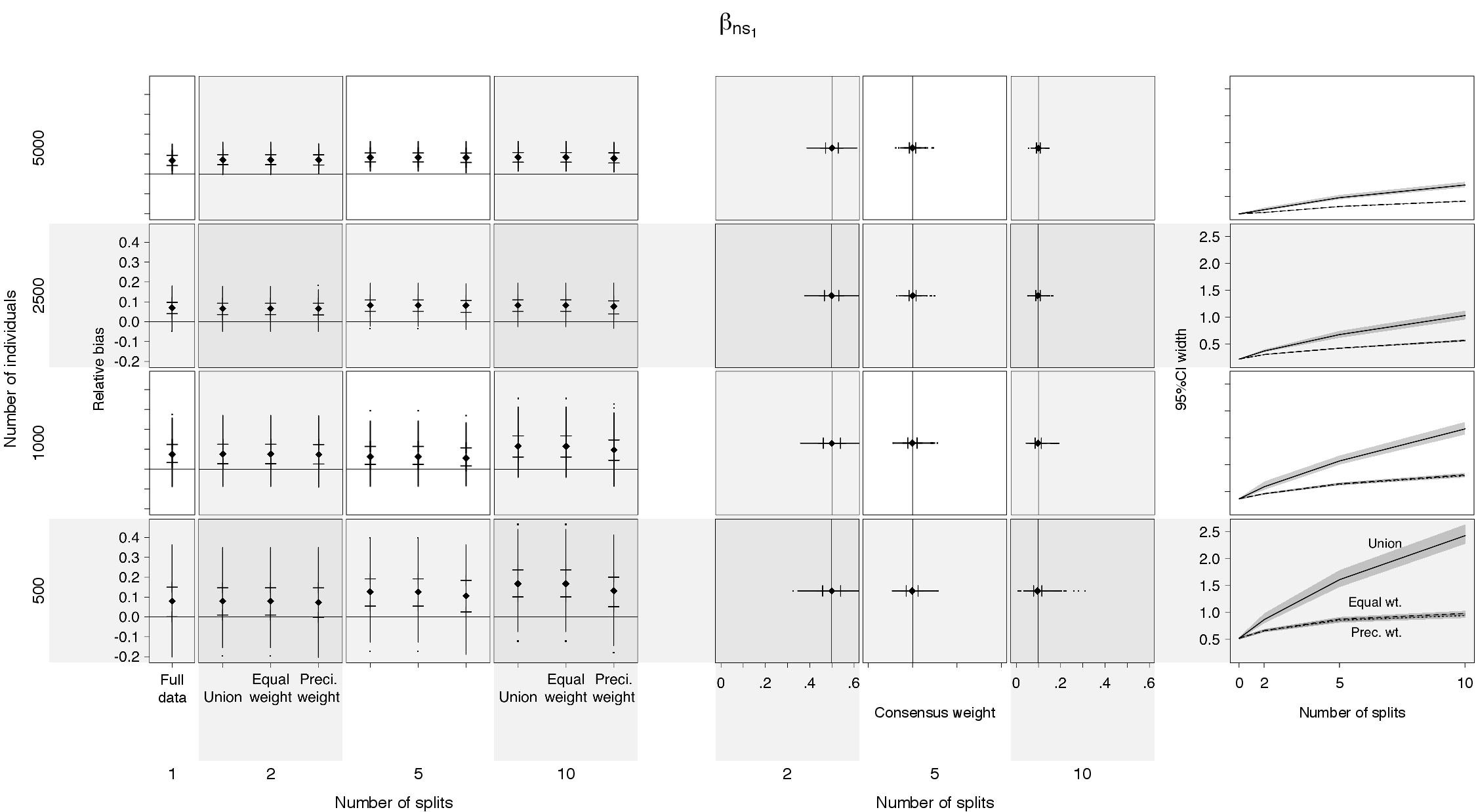}
\caption{Left: Box plot of the relative bias for the $\beta_{\text{ns}_1}$ estimate for different sample sizes and numbers of data splits. Center: Consensus (standardized) precision weights for the $\beta_{\text{ns}_1}$ estimate for different sample sizes and numbers of data splits. The vertical lines represent the equal weights. Right: Median width of the 95\% credible interval, with the associated IQR, of the $\beta_{\text{ns}_1}$ estimate---$1$st, $2$nd, and $3$rd quartiles---against the number of splits for different sample sizes.}
\label{fig:sim:mix:betas2}
\end{figure}

\begin{figure}
\centering
\includegraphics[width=\textwidth]{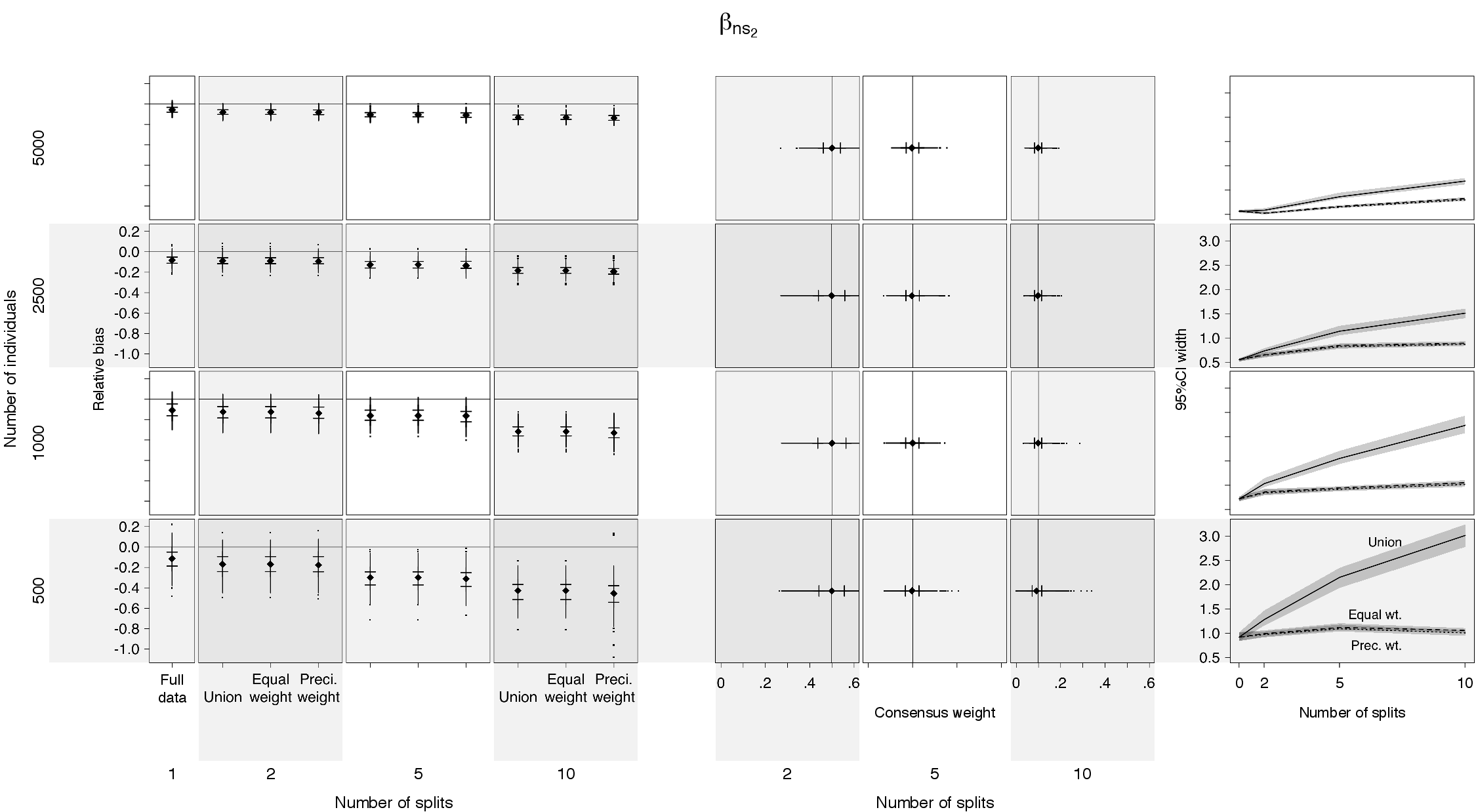}
\caption{Left: Box plot of the  relative bias for $\beta_{\text{ns}_2}$ estimate for different sample sizes and numbers of data splits. Center: Consensus (standardized) precision weights for the $\beta_{\text{ns}_2}$ estimate for different sample sizes and numbers of data splits. The vertical lines represent the equal weight corresponding for that number of data. Right: Median width of the 95\% credible interval, with the associated IQR, of the $\beta_{\text{ns}_2}$ estimate---$1$st, $2$nd, and $3$rd quartiles---against the number of splits for different sample sizes.}
\label{fig:sim:mix:betas3}
\end{figure}

\begin{figure}
\centering
\includegraphics[width=\textwidth]{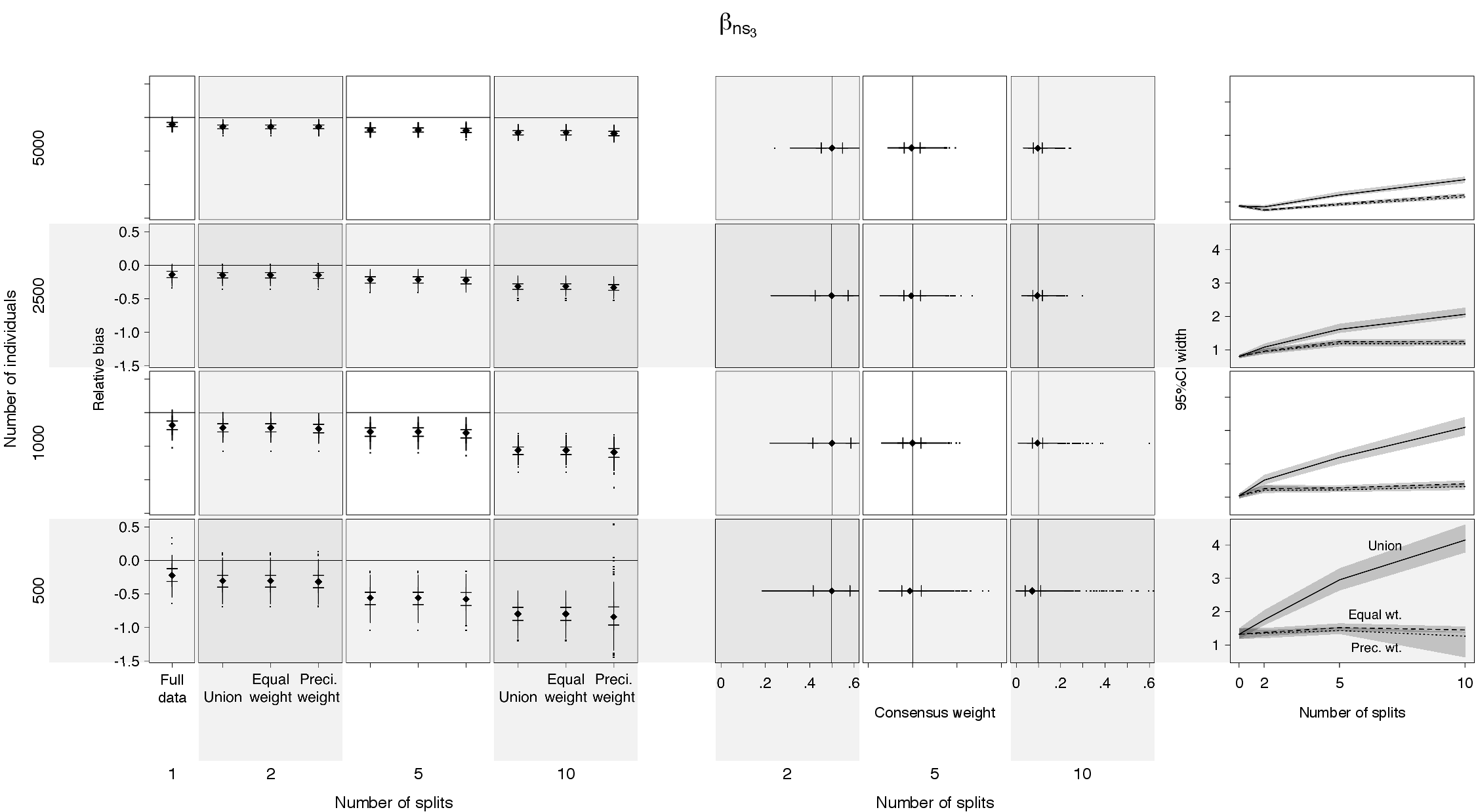}
\caption{Left: Box plot of the  relative bias for $\beta_{\text{ns}_3}$ estimate for different sample sizes and numbers of data splits. Center: Consensus (standardized) precision weights for the $\beta_{\text{ns}_3}$ estimate for different sample sizes and numbers of data splits. The vertical lines represent the equal weight corresponding for that number of data. Right: Median width of the 95\% credible interval, with the associated IQR, of the $\beta_{\text{ns}_2}$ estimate---$1$st, $2$nd, and $3$rd quartiles---against the number of splits for different sample sizes.}
\label{fig:sim:mix:betas4}
\end{figure}

\begin{figure}
\centering
\includegraphics[width=\textwidth]{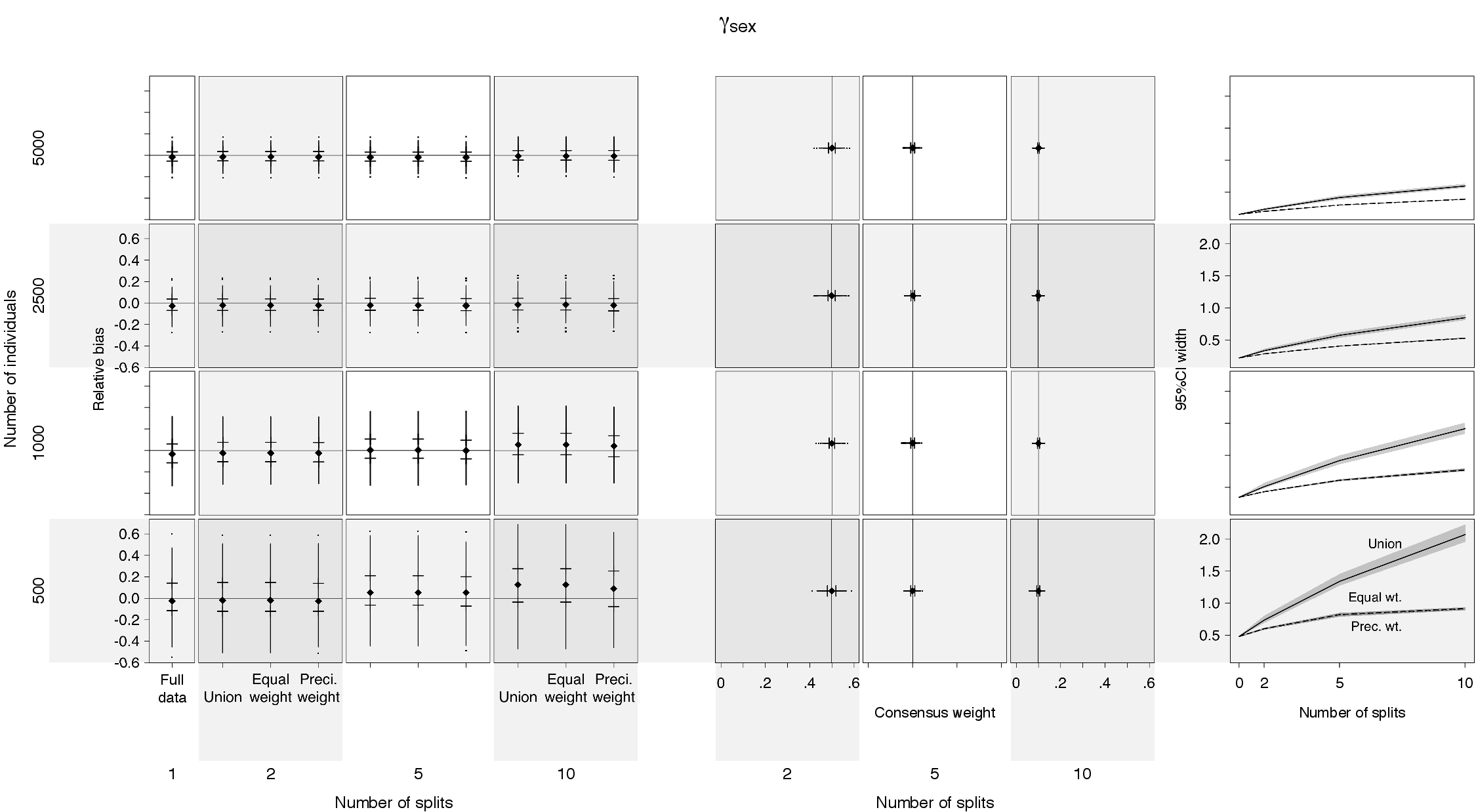}
\caption{Left: Box plot of the relative bias for the $\gamma_{\text{sex}}$ estimate for different sample sizes and numbers of data splits. Center: Consensus (standardized) precision weights for the $\gamma_{\text{sex}}$ estimate for different sample sizes and numbers of data splits. The vertical lines represent the equal weights. Right: Median width of the 95\% credible interval, with the associated IQR, of the $\gamma_{\text{sex}}$ estimate---$1$st, $2$nd, and $3$rd quartiles---against the number of splits for different sample sizes.}
\label{fig:sim:mix:gammas1}
\end{figure}

\begin{figure}
\centering
\includegraphics[width=\textwidth]{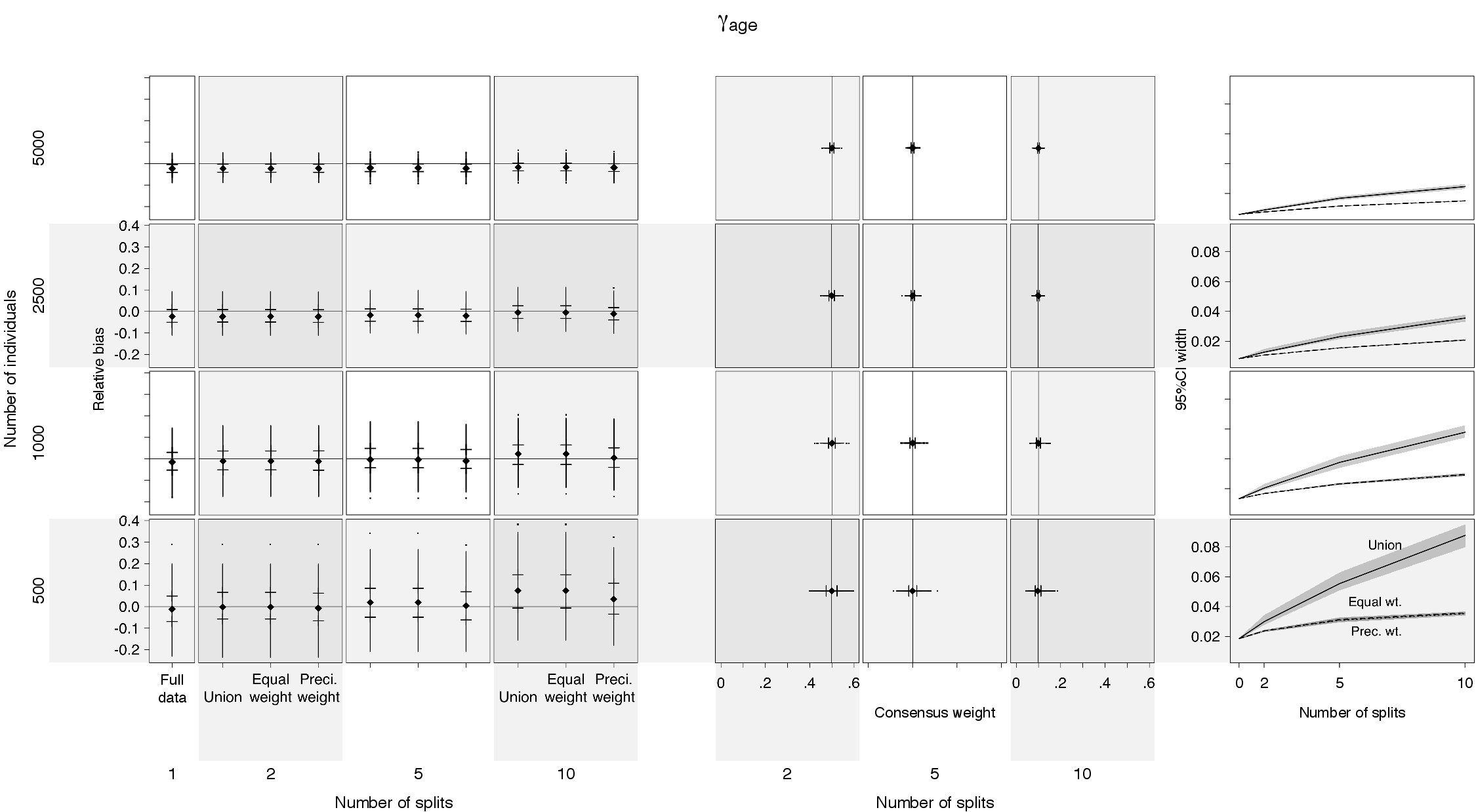}
\caption{Left: Box plot for the relative bias for $\gamma_{\text{age}}$ estimate for different sample sizes and numbers of data splits. Center: Consensus (standardized) precision weights for the $\gamma_{\text{age}}$ estimate for different sample sizes and numbers of data splits. The vertical lines represent the equal weights. Right: Median width of the 95\% credible interval, with the associated IQR, of the $\gamma_{\text{age}}$ estimate---$1$st, $2$nd, and $3$rd quartiles---against the number of splits for different sample sizes.}
\label{fig:sim:mix:gammas2}
\end{figure}

\begin{landscape}

\section{Application study}

\begin{table}[h]
\centering
\caption{Estimated posterior means and 95\% credible interval for the joint model coefficients obtained from the different consensus methods.\label{tab:app:est2}}
\small
\begin{tabular*}{600pt}{@{\extracolsep\fill}lllllll@{\extracolsep\fill}}
\toprule
& \multicolumn{6}{@{}l}{\textbf{Consensus methods}} \\
\cmidrule{2-7}
& \multicolumn{2}{@{}l}{\textbf{Union}} & \multicolumn{2}{@{}l}{\textbf{Equal weight}} & \multicolumn{2}{@{}l}{\textbf{Precision weight}} \\
\cmidrule{2-3} \cmidrule{4-5} \cmidrule{6-7}
\textbf{Param.} & \textbf{Mean}  & \textbf{95\% CI} & \textbf{Mean}  & \textbf{95\% CI} & \textbf{Mean}  & \textbf{95\% CI} \\
\midrule
\textbf{LME} & & & \\
$\beta_0$ & $\hphantom{-0}92.097$ & $\hphantom{-0}(90.012, \hphantom{-0}93.915)$ & $\hphantom{-0}92.097$ & $\hphantom{-0}(91.293, \hphantom{-0}92.913)$ & $\hphantom{-0}92.107$ & $\hphantom{-0}(91.309, \hphantom{-0}92.919)$ \\
$\beta_{\text{ns}_1}$ & ${-288.629}$ & $({-298.948}, {-274.531})$ & ${-288.629}$ & $({-290.853}, {-286.377})$ & ${-288.644}$ & $({-290.858}, {-286.396})$ \\
$\beta_{\text{ns}_2}$ & ${-459.129}$ & $({-480.053}, {-428.993})$ & ${-459.129}$ & $({-463.062}, {-455.213})$ & ${-459.181}$ & $({-463.108}, {-455.269})$ \\
$\beta_\text{pa}$ & $\hphantom{00}{-1.342}$ & $\hphantom{00}({-1.479}, \hphantom{00}{-1.224})$ & $\hphantom{00}{-1.342}$ & $\hphantom{00}({-1.393}, \hphantom{00}{-1.290})$ & $\hphantom{00}{-1.342}$ & $\hphantom{00}({-1.393}, \hphantom{00}{-1.290})$ \\ 
$\beta_{\text{ins}}$ & $\hphantom{00}{-0.154}$ & $\hphantom{00}({-0.361}, \hphantom{-00}0.012)$ & $\hphantom{00}{-0.154}$ & $\hphantom{00}({-0.214}, \hphantom{00}{-0.092})$ & $\hphantom{00}{-0.158}$ & $\hphantom{00}({-0.219}, \hphantom{00}{-0.097})$ \\
$\beta_\text{dpx}$ & $\hphantom{00}{-0.611}$ & $\hphantom{00}({-2.453}, \hphantom{-00}1.540)$ & $\hphantom{00}{-0.611}$ & $\hphantom{00}({-1.116}, \hphantom{00}{-0.090})$ & $\hphantom{00}{-0.632}$ & $\hphantom{00}({-1.136}, \hphantom{00}{-0.112})$ \\ 
$\beta_\text{[88,93)}$ & $\hphantom{00}{-0.376}$ & $\hphantom{00}({-2.513}, \hphantom{-00}1.910)$ & $\hphantom{00}{-0.376}$ & $\hphantom{00}({-1.389}, \hphantom{-00}0.642)$ & $\hphantom{00}{-0.373}$ & $\hphantom{00}({-1.380}, \hphantom{-00}0.640)$ \\ 
$\beta_\text{[93,98)}$ & $\hphantom{00}{-0.045}$ & $\hphantom{00}({-2.056}, \hphantom{-00}2.081)$ & $\hphantom{00}{-0.045}$ & $\hphantom{00}({-1.071}, \hphantom{-00}1.005)$ & $\hphantom{00}{-0.054}$ & $\hphantom{00}({-1.072}, \hphantom{-00}0.991)$ \\ 
$\beta_\text{[98,11]}$ & $\hphantom{-00}3.809$ & $\hphantom{-00}(2.021, \hphantom{-00}5.663)$ & $\hphantom{-00}3.809$ & $\hphantom{-00}(2.993, \hphantom{-00}4.619)$ & $\hphantom{-00}3.826$ & $\hphantom{-00}(3.017, \hphantom{-00}4.629)$ \\
$\beta_\text{sex}$ & $\hphantom{-00}1.529$ & $\hphantom{-00}(0.507, \hphantom{-00}2.570)$ & $\hphantom{-00}1.529$ & $\hphantom{-00}(0.868, \hphantom{-00}2.177)$ & $\hphantom{-00}1.529$ & $\hphantom{-00}(0.868, \hphantom{-00}2.176)$ \\
$\beta_\text{htz}$ & $\hphantom{-00}2.427$ & $\hphantom{-00}(1.073, \hphantom{-00}3.859)$ & $\hphantom{-00}2.427$ & $\hphantom{-00}(1.720, \hphantom{-00}3.145)$ & $\hphantom{-00}2.435$ & $\hphantom{-00}(1.731, \hphantom{-00}3.153)$ \\ 
$\beta_\text{oth}$ & $\hphantom{-00}3.218$ & $\hphantom{-00}(0.807, \hphantom{-00}6.330)$ & $\hphantom{-00}3.218$ & $\hphantom{-00}(2.258, \hphantom{-00}4.183)$ & $\hphantom{-00}3.248$ & $\hphantom{-00}(2.287, \hphantom{-00}4.213)$ \\ 
$\sigma_\text{y}$  & $\hphantom{-00}8.844$ & $\hphantom{-00}(8.805, \hphantom{-00}8.881)$ & $\hphantom{-00}8.844$ & $\hphantom{-00}(8.819, \hphantom{-00}8.863)$ & $\hphantom{-00}8.844$ & $\hphantom{-00}(8.819, \hphantom{-00}8.863)$ \\
\textbf{PH} & & & \\
$\gamma_\text{sex}$ & $\hphantom{00}{-0.425}$ & $\hphantom{00}({-0.628}, \hphantom{00}{-0.252})$ & $\hphantom{00}{-0.425}$ & $\hphantom{00}({-0.519}, \hphantom{00}{-0.332})$ & $\hphantom{00}{-0.420}$ & $\hphantom{00}({-0.514}, \hphantom{00}{-0.328})$ \\
$\alpha$ & $\hphantom{00}{-0.120}$ & $\hphantom{00}({-0.132}, \hphantom{00}{-0.110})$ & $\hphantom{00}{-0.120}$ & $\hphantom{00}({-0.124}, \hphantom{00}{-0.114})$ & $\hphantom{00}{-0.119}$ & $\hphantom{00}({-0.123}, \hphantom{00}{-0.113})$ \\
\bottomrule
\end{tabular*}
\begin{tablenotes}%%[341pt]
\item[] \hspace{0.9cm} CI: credible interval; LME: linear mixed-effects model; PH: proportional hazards model.
\end{tablenotes}
\end{table}
\end{landscape}

\begin{sidewaystable}
\caption{Follow-up, demographic, social, and clinical characteristics of the CF individuals analyzed.\label{tab:app:data}}%
\tiny
\begin{tabular*}{\textheight}{@{\extracolsep\fill}lllllll@{\extracolsep\fill}}%
\toprule
& & \multicolumn{5}{@{}l}{\textbf{Subsamples}} \\
\cmidrule{3-7}
& \textbf{Full data} & \textbf{1} & \textbf{2} & \textbf{3} & \textbf{4} & \textbf{5} \\ 
\midrule
\textbf{Number of individuals} & 35,153 & 7,030 & 7,031 & 7,031 & 7,031 & 7,030\\
\textbf{Birth cohort} & \\
\multicolumn{1}{r}{[1924, 1988)} & 14,473 (41.17\%) & 2,905 (41.32\%) & 2,896 (41.19\%) & 2,925 (41.60\%) & 2,885 (41.03\%) & 2,862 (40.71\%)\\
\multicolumn{1}{r}{[1988, 1993)} & \hphantom{0}4,762 (13.55\%) & \hphantom{0,}934 (13.29\%) & \hphantom{0,}943 (13.41\%) & \hphantom{0,}961 (13.67\%) & \hphantom{0,}967 (13.75\%) & \hphantom{0,}957 (13.61\%)\\
\multicolumn{1}{r}{[1993, 1998)} & \hphantom{0}4,630 (13.17\%) & \hphantom{0,}915 (13.02\%) & \hphantom{0,}948 (13.48\%) & \hphantom{0,}907 (12.90\%) & \hphantom{0,}929 (13.21\%) & \hphantom{0,}931 (13.24\%)\\
\multicolumn{1}{r}{[1998, 2011]} & 11,288 (32.11\%) & 2,276 (32.38\%) & 2,244 (31.91\%) & 2,238 (31.83\%) & 2,250 (32.00\%) & 2,280 (32.43\%)\\
& \\
\textbf{Baseline age (years)} & \\
\multicolumn{1}{r}{Median (IQR)} & 8.92 (6.23, 18.56) & 8.98 (6.23, 18.77) & 9.06 (6.23, 18.49) & 8.93 (6.23, 19.13) & 8.86 (6.23, 18.26)) & 8.77 (6.22, 18.20)\\
& \\
\textbf{Terminal event} & \\
\multicolumn{1}{r}{Censoring} & 26,902 (76.53\%) & 5,383 (76.57\%) & 5,370 (76.38\%) & 5,326 (75.71\%) & 5,373 (76.42\%) & 5,450 (77.53\%)\\
\multicolumn{1}{r}{Death or transplantation} & \hphantom{0}8,251 (23.47\%) & 1,647 (23.43\%) & 1,661 (23.62\%) & 1,705 (24.25\%) & 1,658 (23.58\%) & 1,580 (22.48\%)\\
& \\
\textbf{Age at end of follow-up (years)} & \\
\multicolumn{1}{r}{Censoring, median (IQR)} & 21.33 (14.12, 30.94) & 21.32 (14.04, 31.49) & 21.33 (14.17, 30.63) & 21.33 (14.06, 30.86) & 21.43 (14.20, 30.88) & 21.30 (14.12, 30.96)\\
\multicolumn{1}{r}{Death/transplantation, median (IQR)} & 27.12 (21.36, 35.99) & 26.69 (21.20, 36.16) & 27.61 (21.52, 36.44) & 27.47 (21.55, 36.20) & 26.82 (21.16, 35.38) & 27.10 (21.33, 35.44)\\
& \\
\textbf{Genotype (F508del)} & \\
\multicolumn{1}{r}{Homozygous} & 15,656 (44.54\%) & 3,116 (44.32\%) & 3,130 (44.52\%) & 3,165 (45.02\%) & 3,114 (44.29\%) & 3,131 (44.54\%)\\
\multicolumn{1}{r}{Heterozygous} & 14,002 (39.83\%) & 2,820 (40.11\%) & 2,774 (39.45\%) & 2,790 (39.68\%) & 2,815 (40.04\%) & 2,803 (39.87\%)\\
\multicolumn{1}{r}{Neither} & \hphantom{0}5,495 (15.63\%) & 1,094 (15.53\%) & 1,127 (16.03\%) & 1,076 (15.30\%) & 1,102 (15.67\%) & 1,096 (15.59\%)\\
& \\
\textbf{Sex} & \\
\multicolumn{1}{r}{Female} & 16,992 (48.34\%) & 3,392 (48.25\%) & 3,375 (48.00\%) & 3,378 (48.04\%) & 3,376 (48.02\%) & 3,471 (49.37\%)\\
& \\
\textbf{Number of ppFEV\textsubscript{1} measurements} & 1,523,406 & 304,119 & 305,358 & 302,592 & 304,541 & 306,796\\
& \\
\textbf{Number of ppFEV\textsubscript{1} measurements/ind.} & \\
\multicolumn{1}{r}{Median (IQR)} & 36.00 (15.00, 64.00) & 36.00 (15.00, 63.75) & 36.00 (15.00, 64.00) & 36.00 (15.00, 62.00) & 36.00 (15.00, 64.00) & 36.00 (16.00, 63.00)\\
& \\
\textbf{Total follow-up duration (years)} & 372,365.79 & 74,084.40 & 74,478.90 & 74,411.97 & 74,393.33 & 74,997.19 \\
& \\
\textbf{Follow-up duration/ind. (years)} & \\
\multicolumn{1}{r}{Median (IQR)} & 10.28 (4.59, 16.78) & 10.26 (4.47, 16.70) & 10.33 (4.50, 16.78) & 10.16 (4.64, 16.79) & 10.19 (4.62, 16.72) & 10.43 (4.71, 16.90)\\
& \\
\textbf{Baseline ppFEV\textsubscript{1}} & \\
\multicolumn{1}{r}{Median (IQR)} & 86.00 (65.50, 100.70) & 86.10 (66.40, 100.80) & 86.40 (65.50, 100.80) & 85.40 (64.90, 100.25) & 86.30 (65.20, 100.80) & 85.70 (65.60, 100.90)\\
& \\
\textbf{Medicaid insurance possession} & \\
\multicolumn{1}{r}{At baseline} & 16,350 (46.51\%) & 3,318 (47.20\%) & 3,233 (45.98\%) & 3,240 (46.08\%) & 3,259 (46.35\%) & 3,300 (46.94\%)\\
\multicolumn{1}{r}{Throughout follow-up} & \hphantom{0}8,677 (24.68\%) & 1,778 (25.29\%) & 1,698 (24.15\%) & 1,708 (24.29\%) & 1,770 (25.17\%) & 1,723 (24.51\%)\\
\multicolumn{1}{r}{Sometime during follow-up} & 28,632 (81.45\%) & 5,707 (81.18\%) & 5,714 (81.27\%) & 5,692 (80.96\%) & 5,727 (81.45\%) & 5,792 (82.39\%)\\
& \\
\textbf{\textit{Pseudomonas aeruginosa}} & \\
\multicolumn{1}{r}{At baseline} & \hphantom{0}3,621 (10.30\%) & \hphantom{0,}769 (10.94\%) & \hphantom{0,}695 \hphantom{0}(9.89\%) & \hphantom{0,}735 (10.45\%) & \hphantom{0,}704 (10.01\%) & \hphantom{0,}718 (10.21\%)\\
\multicolumn{1}{r}{Throughout follow-up} & \hphantom{00,}235 \hphantom{0}(0.67\%) & \hphantom{0,0}66 \hphantom{0}(0.94\%) & \hphantom{0,0}48 \hphantom{0}(0.68\%) & \hphantom{0,0}43 \hphantom{0}(0.61\%) & \hphantom{0,0}36 \hphantom{0}(0.51\%) & \hphantom{0,0}42 \hphantom{0}(0.60\%)\\
\multicolumn{1}{r}{Sometime during follow-up} & 25,970 (73.88\%) & 5,208 (74.08\%) & 5,162 (73.42\%) & 5,181 (73.69\%) & 5,171 (73.55\%) & 5,248 (74.65\%)\\
\bottomrule
\end{tabular*}
\begin{tablenotes}%%[\textheight]
\item[] IQR: interquartile range. 
\end{tablenotes}
\end{sidewaystable}

\end{document}